\documentclass[12pt]{article}
\pdfoutput=1
\usepackage[utf8]{inputenc}
\usepackage[english]{babel}
\usepackage{amsmath}
\usepackage{graphicx}
\graphicspath{ {images/} }
\usepackage{float}
\usepackage{amssymb}
\usepackage{caption}
\usepackage{subcaption}
\usepackage{adjustbox}
\usepackage{tabularx}
\usepackage{multicol}
\usepackage{multirow}
\usepackage{slashed}
\usepackage{feynmp-auto}
\usepackage{comment}

\usepackage[tableposition=below]{caption}
\usepackage{longtable}
\usepackage{color}			
\usepackage{verbatim}

\newcommand{\LL}{\mathcal{L}}

\newcommand{\NN}{\mathcal{N}}

\newcommand{\arXiv}[2]{\href{http://arxiv.org/pdf/#1}{{\tt #2/#1}}}
\newcommand{\arXivold}[1]{\href{http://arxiv.org/pdf/#1}{{\tt #1}}}

\newcommand{\beq}{\begin{eqnarray}}
\newcommand{\eeq}{\end{eqnarray}}

\numberwithin{equation}{section} 
\usepackage[
	colorlinks=true,
	citecolor=red,
	linkcolor=blue,
	urlcolor=blue,
	hypertexnames=false]{hyperref}

\begin{document}
\begin{titlepage}
\vspace{1cm}
\begin{center}
		\LARGE \bf 
		Collider Phenomenology of\\
			\vskip .3cm
		\LARGE \bf a Gluino Continuum
			
\end{center}
	\vskip .3cm
	
	\renewcommand*{\thefootnote}{\fnsymbol{footnote}}

\vspace{0.9cm}
\begin{center}
		
		\bf
		Christina Gao$^{a,}$\footnote{\tt \scriptsize
		 \href{mailto:yanggao@fnal.gov}{yanggao@fnal.gov},
		 $^\dag$\href{mailto:shayegan@ucdavis.edu}{shayegan@ucdavis.edu}, $^\ddagger$\href{mailto:jterning@gmail.com}{jterning@gmail.com}
		 },
		Ali Shayegan Shirazi$^{b,\dag}$, and John Terning$^{b,\ddagger}$
\end{center}
	
	\renewcommand{\thefootnote}{\arabic{footnote}}
	\setcounter{footnote}{0}


\begin{center} 

 	{\it a Theoretical Physics Department, Fermilab, Batavia, IL 60510}\\
	{\it b Center for Quantum Mathematics and Physics (QMAP)\\ Department of Physics, University of California\\Davis CA 95616}

\end{center}

\vspace{1cm}

\centerline{\large\bf Abstract}
\begin{quote}
Continuum supersymmetry is a class of models in which the supersymmetric partners together with part of the standard model  come from a conformal sector, broken in the IR near the TeV scale. Such models not only open new doors for addressing the problems of the standard model, but also have unique signatures at hadron colliders, which might explain why we have not yet seen any superpartners at the LHC. Here we use gauge-gravity duality to model the conformal sector, generate collider simulations, and finally analyze continuum gluino signatures at the LHC. Due to the increase in the number of jets produced the bounds are weaker than for the minimal supersymmetric standard model with the same gluino mass threshold.
 \end{quote}

\end{titlepage}

\section{Introduction}

Despite years of searches for superpartners at the LHC, there have only been rising lower bounds rather than discoveries.  The constraints on gluinos are particularly strong since they have a large production cross section, however very large gluino masses lead to squark instabilities, the breaking of color, and a complete failure of phenomenology.
Media pundits bewail the lack of imagination required to press on with superpartner searches, but we believe it is completely necessary to push the existing searches as far as they can go.  On the other hand, it is also valuable to explore non-minimal supersymmetric scenarios where the current searches may not be adequate.  One interesting possibility is that the supersymmetric sector is part of a strongly coupled sector.  There are many possibilities along these lines, but the simplest variation to explore phenomenologically is when the strong sector approaches an IR fixed point, and some approximate conformal symmetry is manifest.

The idea that a conformal sector could manifest itself with collider signatures was first made plausible by Georgi \cite{georgi}. The conformal ``stuff" that one  could detect in a collider was named \emph{unparticles}. The unparticles either weakly couple to the standard model (SM) \cite{Fox:2007sy, Cacciapaglia:2007jq, Delgado:2007dx, Kikuchi:2007qd, Delgado:2008rq, Delgado:2008px, Lee:2008ph, Espinosa:1998xj}, or SM particles, eg. the Higgs, are themselves part of the conformal sector \cite{Stancato:2008mp, Bellazzini:2015cgj}. The basic signature is a continuum of particles; more or less like a spectrum of particle masses too closely spaced to be individually resolved \cite{vanderBij:2007um, Englert:2012dq, Goncalves:2018pkt, Csaki:2018kxb, Megias:2019vdb, Shirazi:2019bjw}. For the  SM particles---or their supersymmetric partners---to be unparticles, the conformal sector should be broken at not-too-low an energy. A phenomenologically plausible continuum cannot extend all the way to the true IR since this corresponds to new massless degrees of freedom. To preserve a conformal/unparticle signature above the breaking scale, the conformal symmetry must be broken \emph{softly} \cite{Cacciapaglia:2007jq}. We will use the symbol $\mu$ for the conformal breaking scale, this scale acts as a threshold for the production of the unparticle ``stuff."

While it is hard to deal with strongly coupled conformal field theories (CFT's) directly, we can alternatively work in a 5D anti-de Sitter (ADS) spacetime using the AdS/CFT correspondence \cite{Maldacena:1997re, ArkaniHamed:2000ds} to model such theories. According to this conjecture, fields that are part of the 4D conformal sector correspond to propagating degrees of freedom in a 5D spacetime with an asymptotically AdS background, while the isometries of the AdS space correspond to the conformal symmetries of the 4D theory. In the 5D picture, the soft breaking is achieved by a \emph{soft wall} \cite{08terning2, Falkowski:2008yr, Batell:2008me}. A soft wall, roughly, is a field that has a nonzero vacuum expectation value (VEV) which depends on the fifth coordinate. As a result, the background geometry away from the AdS boundary is deformed, hence the conformal symmetry in the dual gauge theory is softly-broken in the IR. This is in contrast to Randall-Sundrum (RS) models \cite{RS,Rattazzi:2000hs}, where the breaking is done by a hard wall, i.e. an IR brane, which leads to Kaluza-Klein (KK) resonances in the 4D theory. 

We will assume that the two lightest generations of quarks, as well as the gauge bosons, and their superpartners, propagate in the 5D bulk. Each supermultiplet comes with their own soft wall and hence breaking (threshold) scale $\mu_i$. Because of SUSY, the threshold scale for each field and its supersymmetric partner are the same. The rest of the minimal supersymmetric standard model (MSSM) is assumed to live on a UV brane, a 4D boundary of the 5D spacetime.

 Without a SUSY breaking mechanism, the spectrum of each of the fields is described by a zero mode and a continuum that starts at $\mu_i$. To break SUSY, one can add 4D mass terms on the UV brane, with a scale $\sim M_{SUSY}$, so that the zero modes of the superpartners are no longer degenerate with the SM fields. In particular, by varying the SUSY breaking mass, the zero mode of the superpartner can be lifted until it merges with the continuum.

To explore the collider phenomenology of these conformal/unparticle superpartners, it is easiest to approximate them by towers of closely spaced particles \cite{ArkaniHamed:2001ca,Stephanov:2007ry}. In theories with AdS duals, these particles are KK modes. The spacing of the KK modes is determined by the position of an IR regulator brane. If the regulator brane is taken to infinity, the KK modes merge to form a continuum. In the KK approximation, we deal with normal particles that are produced and subsequently decay. The production and decay amplitudes of the individual KK particles will depend directly on the location of the IR brane. Since the IR brane is just a regulator, we need to suitably sum over KK modes to find regulator independent quantities.

In particular, we are interested in the production and the decay of continuum gluinos. Once a pair of highly excited KK gluinos are produced, each of them can decay to a quark and a KK squark. Now if the KK squark is heavier than the lightest KK gluino, it can further decay to a second KK gluino and a quark. This gives us a chain decay until the lightest KK gluino or squark is produced. In this work, we will assume that the zero mode of a neutral SUSY-ino is the lightest supersymmetric particle (LSP) and that it has a mass below the continuum mass gap of the KK squarks and gluinos, such that the lightest KK squark (gluino) at the end of the decay chain can further decay into SM quark(s) and the LSP. At the detector level, therefore, the typical signature is jets with high multiplicities and missing transverse energy.

In section \ref{adssusy} we will lay out the necessary  5D field theory technology and discuss how one obtains the effective 4D theory. In section \ref{prodanddecay} we analyze the production and decay of gluino KK modes. In section \ref{lhc} we will discuss the signatures of the gluino continuum that one would expect to see at the LHC. Finally, in section \ref{conclusion} we present our  outlook for the continuum supersymmetry and discuss possible applications to other models.  In Appendix. \ref{app:holoAction} we give a brief review of the AdS/CFT correspondence and in Appendix \ref{app:KKvsCon} we discuss the relation between the regulated KK tower and its continuum limit.


\section{Review of Super-Multiplets in AdS$_5$}\label{adssusy}
 Through the AdS/CFT correspondence\cite{Maldacena:1997re}, we can model a 4D conformal theory using a 5D AdS spacetime (for more details see Appendix. \ref{app:holoAction}). The 5D metric is
\begin{equation}
ds^2=\left(\frac{R}{z}\right)^2(\eta_{\mu\nu}dx^{\mu}dx^{\nu}-dz^2)~,
\end{equation}
where $R$ is the curvature radius, $z$ the fifth coordinate, and $\eta_{\mu\nu}$ the flat Minkowski metric with signature $(+ - - - )$. The space is cut off by a UV brane, which for simplicity we will set at $z_{UV}=R$. 

Since 5D fermions cannot be chiral, a 5D $\mathcal{N}=1$ supermultiplet necessarily contains two  4D chiral superfields 
$\Phi=\{\phi,\chi,F\}$ and $\Phi_c=\{\phi_c,\psi,F_c\}$, where the Weyl fermions $\chi$ and $\psi$ form a 5D Dirac fermion \cite{01pomarol}. 

For the matter fields, the bulk 5D AdS action takes the form \cite{08terning}
\begin{equation}
\begin{split}
S_{matter}=&\int d^4x dz \Big\{\int d^2\theta d^2\bar{\theta} \left(\frac R z \right)^3 [\Phi^*\Phi+\Phi_c\Phi_c^*]\\
&+\int d^2\theta\left(\frac R z \right)^3 [\frac 1 2\Phi_c\partial_z \Phi-\frac12 \partial_z\Phi_c\Phi+m(z)\frac Rz \Phi_c\Phi]+h.c.\Big\}
\end{split}
\end{equation}
where  $m(z)R=c+\mu z$.  Through the AdS/CFT dictionary, $c$ is related to the scaling dimension of the dual fields. For the left-handed (LH) fermion, $\chi$, 
we have $d_f=2-c$, and for the scalar $\phi$ 
, $d_s=3/2-c$. For the right-handed (RH) fermion $\psi$ and $\phi_c$,   the scaling dimensions are given by $c\to-c$. Note that the relation $d_f=d_s+1/2$ is required if these fields form a superconformal multiplet \cite{Minwalla:1997ka}. 

The $z$-dependent part of the mass, $\mu z$,  generates a mass gap for the continuum \cite{09terning}. It violates 5D Lorentz invariance and half of the SUSY but the 4D Lorentz Invariance and 4D $\mathcal{N}=1$ SUSY is preserved. 

Expanding the superfields one gets
\begin{equation}
\begin{split}\label{matteraction}
S_{matter}=&\int d^4 x \int dz  \left(\frac R z\right)^3 
\Big\{\\
&
[ \partial_{\mu}\phi\partial^{\mu}\phi^*-i\bar{\chi}\bar{\sigma}^{\mu}\partial_{\mu}\chi+FF^*+ \partial_{\mu}\phi_c\partial^{\mu}\phi^*_c-i\bar{\psi}\bar{\sigma}^{\mu}\partial_{\mu}\psi+F_cF_c^*]\\
&+[\frac 12 (\phi_c\partial_z F+F_c\partial_z \phi-\chi\partial_z\psi-F\partial_z\phi_c-\phi\partial_zF_c+\psi\partial_z\chi)\\
&+\frac{m(z)R}{z}(\phi_cF+F_c\phi+\chi\psi)+h.c.]
\Big\}
\end{split}
\end{equation}
where the lowering and raising of the index is with respect to the metric $\eta_{\mu\nu}$.

By varying the bulk action with respect to each field, one can obtain their equations of motion. The bulk fields are decomposed as a product of a 4D field and a 5D profile (the LH and RH fermions $\chi,\psi$ in (\ref{matteraction}) have been rescaled by $( z/ R)^{1/2}$) \cite{09terning}:
\begin{equation}\label{rhfield}
\begin{split}
\chi(p,z)&=\chi_4(p)\left(\frac{z}{z_{UV}}\right)^2f_L(p,z)~, 
\phi(p,z)=\phi_4(p)\left(\frac{z}{z_{UV}}\right)^{3/2}f_L(p,z)~,
\\
\psi(p,z)&=\psi_4(p)\left(\frac{z}{z_{UV}}\right)^2f_R(p,z)~,
\phi_c(p,z)=\phi_{c4}(p)\left(\frac{z}{z_{UV}}\right)^{3/2}f_R(p,z)~,
\end{split}
\end{equation}
where $\chi_4,\psi_4$ solve the 4D Dirac equation,
\begin{equation}
-i\bar{\sigma}^{\mu}\partial_{\mu}\chi_4+p\bar{\psi}_4=0,\quad-i\sigma^{\mu}\partial_{\mu}\bar{\psi}_4+p\chi_4=0.
\end{equation}
In terms of the 5D profiles, the equation of motion reads:
\begin{equation}\label{1stordereom}
p f_L+\left(\partial_z-\frac{m(z)R}{z}\right)f_R=0~,\quad p f_R-\left(\partial_z+\frac{m(z)R}{z}\right)f_L=0~.
\end{equation}
The normalization of the fields is given by
\begin{equation}\label{eq:normal}
\int_{z_{UV}}^{z_{IR}}dz|f_L|^2=\int _{z_{UV}}^{z_{IR}}dz|f_R|^2=1.
\end{equation}
 For convenience we will denote the normalization factors by $\NN_{L,R}$ such that 
\begin{equation}\label{solutions2}
\begin{split}
f_L(p,z)=&\NN_L(p)\mathfrak{f}_L(p,z)~,\\
f_R(p,z)=&\NN_R(p)\mathfrak{f}_R(p,z)~.\\
\end{split}
\end{equation}
We will make the $\mathfrak{f}$'s dimensionless, and, from \eqref{eq:normal}, the dimension of $\NN_{L,R}$ is $1/2$. 

For the zero modes, $p=0$, there exist solutions of the form
\begin{equation}\label{0mode}
\mathfrak{f}^0_L\sim e^{-\mu z}\left(\frac{z}{z_{UV}}\right)^{-c},\quad \mathfrak{f}^0_R\sim e^{\mu z}\left(\frac{z}{z_{UV}}\right)^c.
\end{equation}
The normalization of the LH zero modes is then given by
\begin{equation}\label{eq:normzeromode}
\NN_L(0)=\Big(2^{-1+2c}\mu^{-1}(z_{UV}\mu)^{2c}\Gamma(1-2c)\Big)^{-1/2}~.
\end{equation}
Note that for $\mu>0$, only $\mathfrak{f}_L^0$ is finite in the IR. Later on, we get rid of the RH zero mode by imposing the boundary condition $\Phi_c|$=0, so that the LH  zero mode can be interpreted as the SM fields.

For non-zero mode solutions, we combine \eqref{1stordereom} and find
\beq\label{eomfl}
\left[\partial_z^2+p^2-\mu^2-\frac{2\mu c}{z}-\frac{c(c+1)}{z^2}\right]f_L&=0~,\\
\left[\partial_z^2+p^2-\mu^2-\frac{2\mu c}{z}-\frac{c(c-1)}{z^2}\right]f_R&=0~.
\eeq
The solutions are linear combinations of Whittaker functions \cite{Abramowitz:1965},
\begin{equation}\label{solutions}
\begin{split}
f_L(p,z)=&A
M(\kappa,\frac12+c,2\sqrt{\mu^2-p^2}z)+BW(\kappa,\frac12+c,2\sqrt{\mu^2-p^2}z)~,\\
f_R(p,z)=&-A\frac{2c(1+2c)\sqrt{\mu^2-p^2}}{p}
M(\kappa,-\frac12+c,2\sqrt{\mu^2-p^2}z)\\
&-B\frac{p}{\mu+\sqrt{\mu^2-p^2}}W(\kappa,-\frac12+c,2\sqrt{\mu^2-p^2}z)~,
\end{split}
\end{equation}
where
\beq
 \kappa\equiv \frac{-c\mu}{\sqrt{\mu^2-p^2}}~.
\eeq
In terms of $\mathfrak{f}_L$ and $\mathfrak{f}_R$ \eqref{solutions2}, we have
\begin{equation}\label{solutions3}
\begin{split}
\mathfrak{f}_L(p,z)=&M(\kappa,\frac12+c,2\sqrt{\mu^2-p^2}z)+\mathbf{b}(p)W(\kappa,\frac12+c,2\sqrt{\mu^2-p^2}z)~,\\
\mathfrak{f}_R(p,z)=-&\frac{2c(1+2c)\sqrt{\mu^2-p^2}}{p}
M(\kappa,-\frac12+c,2\sqrt{\mu^2-p^2}z)\\&-\mathbf{b}(p)\frac{p}{\mu+\sqrt{\mu^2-p^2}}W(\kappa,-\frac12+c,2\sqrt{\mu^2-p^2}z)~.\\
\end{split}
\end{equation}

The boundary condition at the UV brane fixes the remaining coefficient. As mentioned above, we eliminate the RH zero modes by imposing a Dirichlet boundary condition for the RH fields at this brane. This gives 
\begin{equation}
\mathbf{b}(p)=-\frac{M(\kappa,-\frac12+c,2\sqrt{\mu^2-p^2}z_{UV})}{W(\kappa,-\frac12+c,2\sqrt{\mu^2-p^2}z_{UV})}\frac{2(1+2c)\sqrt{\mu^2-p^2}}{\mu-\sqrt{\mu^2-p^2}}~.
\end{equation}
%


\subsection{Gauge Fields}
In the gauge sector, a 5D $\mathcal{N}=1$ vector superfield can be decomposed into a 4D vector superfield $V=\{A_{\mu},\lambda_1,D\}$ and a 4D chiral superfield $\sigma=\{(\Sigma+iA_5)/\sqrt{2},\lambda_2,F_{\sigma}\}$\cite{01pomarol}. 
Because of 5D Gauge invariance, a $z-$dependent bulk mass can no longer be introduced to break the CFT. One way to achieve a similar effect is to introduce a dilaton interaction in the bulk~\cite{09terning}:
\begin{equation}
\begin{split}
S_{A}=&\int d^4x\,dz \frac Rz 
\Big\{\frac14\int d^2\theta  W_{\alpha}W^{\alpha}\Phi +h.c.\\
&+\frac12\int d^4\theta \Big(\partial_zV-\frac Rz\frac{(\sigma+\sigma^{\dagger})}{\sqrt{2}}\Big)^2(\Phi+\Phi^{\dagger})
\Big\}~,
\end{split}
\end{equation}
 where the scalar component of the dilaton field $\Phi$ acquires a VEV $\langle\Phi\rangle=\frac{e^{-2\mu z}}{g_5^2}$.

After expanding the superfields in Wess-Zumino gauge, we find
 \begin{equation}\label{forceaction}
 \begin{split}
 S_{A}=&\int d^4xdz\frac Rz e^{-2\mu z}\Big\{\\
 &
 -\frac 14 F^a_{\mu\nu}F^{\mu\nu a}+i\lambda_1^{\dagger a}\bar{\sigma}^{\mu}D_{\mu}\lambda_1^a+\frac12D^aD^a -\frac12 (\partial_zA_{\mu})^2\\
 &+
 \left(\frac Rz\right)^2 \big [
 \frac12(\partial_{\mu}\Sigma)^2-\frac12(\partial_{\mu}A_5)^2+i\lambda_2^{\dagger}\bar{\sigma}^{\mu}\partial_{\mu}\lambda_2 + F_{\sigma}^*F_{\sigma}  \big]\\
 &+
 \left(\frac Rz\right)
 \big[
 \lambda_2\partial_z\lambda_1+\lambda_2^{\dagger}\partial_z\lambda_1^{\dagger}-\partial_zA_{\mu}\partial^{\mu}A_5
 -\Sigma\partial_zD
 \big]
 \Big\}~,
 \end{split}
 \end{equation}
 where  $D_{\mu}\lambda^a_1=\partial_{\mu}\lambda^a_1-g_5 f^{abc}A_{\mu}^b\lambda_1^c$, and $g_5$ is the 5D gauge coupling, with mass dimension $-1/2$.
 
As shown in \cite{09terning}, the solutions to the bulk equations of motion can be decomposed into a 4D field multiplied by the $z-$dependent profile\footnote{$\lambda_2$ has been rescaled by $(\frac z R)^{1/2}$ and $\lambda_1$ has been rescaled by $(\frac z R)^{3/2}$},
\begin{equation}\label{gaugefields}
\begin{split}
\lambda_1(p,z)=\lambda_4(p)e^{\mu z}\left(\frac{z}{z_{UV}}\right)^{2}h_L(p,z)~, \quad
&A_{\mu}(p,z)=A_{\mu4}(p)e^{\mu z}\left(\frac{z}{z_{UV}}\right)^{1/2}h_L(p,z)~,\\
\lambda_2(p,z)=\lambda'_4(p)e^{\mu z}\left(\frac{z}{z_{UV}}\right)^{2}h_R(p,z)~,\quad
&\Sigma(p,z)=\Sigma_{4}(p)e^{\mu z}\left(\frac{z}{z_{UV}}\right)^{3/2}h_R(p,z)~,
\end{split}
\end{equation}
where $h_{L,R}$ are simply $f_{L,R}$ \eqref{solutions} evaluated at $c=1/2$. We again define
\beq
h_L(p,z)=\NN_L(p)\mathfrak{h}_L(p,z)~,\\
h_R(p,z)=\NN_R(p)\mathfrak{h}_R(p,z)~,
\eeq
 where $\mathfrak{h}_{L,R}$ are  $\mathfrak{f}_{L,R}$ \eqref{solutions3} evaluated at $c=1/2$. The normalization is  found by imposing
\begin{equation}\label{normalization}
\int_{z_{UV}}^{z_{IR}} dz |h_L|^2 =1~.
\end{equation}
One can for example check that the gauge boson zero mode 
\beq
A_{\mu}^0=A^0_{\mu4}(p)e^{\mu z}\left(\frac{z}{z_{UV}}\right)^{1/2}h^0_L(p,z)~,
\eeq
is indeed flat as required by gauge invariance.


\subsection{SUSY Breaking}
In the real world, supersymmetry must be broken, so we must preserve the zero modes of the chiral fermions (which are the SM quarks and leptons) while giving masses to lift the zero mode scalars in order to split the supermultiplet. Following the discussion in \cite{09terning}, for the matter fields, we introduce SUSY breaking on the UV boundary by a scalar mass term:
\begin{equation}
\delta S=\frac12 \int d^4x dz \left(\frac Rz \right)^3 (M_{SUSY}^2z_{UV}\phi^*\phi+ h.c.)\delta(z-z_{UV})~.
\end{equation}
This modifies the boundary condition for the auxillary field $F$:
\begin{equation}
F_c(z_{UV})=M_{SUSY}^2z_{UV}\phi^*(z_{UV})~,
\end{equation}
and therefore,
\begin{equation}\label{modifiedUV}
pf_R(p,z_{UV})-M_{SUSY}^2z_{UV}f_L(p,z_{UV})=0~.
\end{equation}

 Given $M_{SUSY}$, the modified UV boundary condition together with the Dirichlet IR boundary condition for the RH fields determine the mass of the lifted zero mode.

In order to mimic the behavior of an unparticle, we can lift the mass of the zero mode anywhere above $\mu_f$, by tuning the SUSY breaking mass. However, for operational simplicity, we will choose the case where the lifted zero mode acquires a mass just below the threshold $\mu_f$. That is, we  take $p\approx \mu_f-\epsilon$, where $\epsilon \ll \mu_f$. For $\mu_f=1$ TeV,  $c=0.45$, $z_{UV}=10^{-3}$ TeV$^{-1}$, we obtain $M_{SUSY}\approx 2.8$ TeV.

To break SUSY for gauge supermultiplets, we need to add a Majorana mass on the UV brane~\cite{09terning},
\begin{equation}
\delta S=\int d^4x dz \left(\frac R {z_{UV}}\right)^4 (M'_{SUSY}\lambda_1\lambda_1+h.c.)\delta(z-z_{UV})~.
\end{equation}
This modifies the UV boundary condition for the gauginos,
\begin{equation}\label{uvbc}
h_R(p,z_{UV})=M'_{SUSY}z_{UV}h_L(p,z_{UV})~.
\end{equation}
For the gluino, we again choose a breaking mass such that the zero mode gluino merges with the continuum. As an example, for $\mu=1$ TeV, $c=$1/2, $z_{UV}=10^{-3}$ TeV$^{-1}$, we have $M'_{SUSY}=5.7$ TeV. 

The required SUSY breaking mass depends on the scaling dimension of the fields through $c$. For $0<c<1/2$, $M_{SUSY}$ increases as $c$ decreases. Since we need $c=1/2$ for the gauge supermultiplet, in order for $M_{SUSY}$ in the matter sector to be of the same order as $M'_{SUSY}$ in the gauge sector, we choose $c\approx 1/2$ in the matter sector. The benchmark value that we have used in all of our plots and simulations is $c=0.45$.


\subsection{Lightest Supersymmetric  Particle}
For this work we assume that the LSP is the lifted bino zero mode, which takes a similar form to the gluino zero mode: 
\begin{equation}\label{binoprofile}
\tilde{B}^0(m_{\chi},z)=\tilde{\chi} \,e^{\mu_b z} \left(\frac zR\right)^{2} h^0_{L}(m_{\chi},z)~.
\end{equation} 
where up to normalization constant,
\begin{equation}\label{bino}
h_L^0(p,z) = e^{-z\sqrt{\mu_b^2-p^2}}\Big(z\sqrt{\mu_b^2-p^2}\Big)^{1+c}U\Big(1+c-\kappa,2+2c,2z\sqrt{\mu_b^2-p^2}\Big)~
\end{equation}
 The 4D field, which we have called $\tilde{\chi}$, is the LSP, while  $\mu_b$ is the threshold scale for the hypercharge sector.
 
\begin{figure}[ht!]
\centering
\includegraphics[scale=.7]{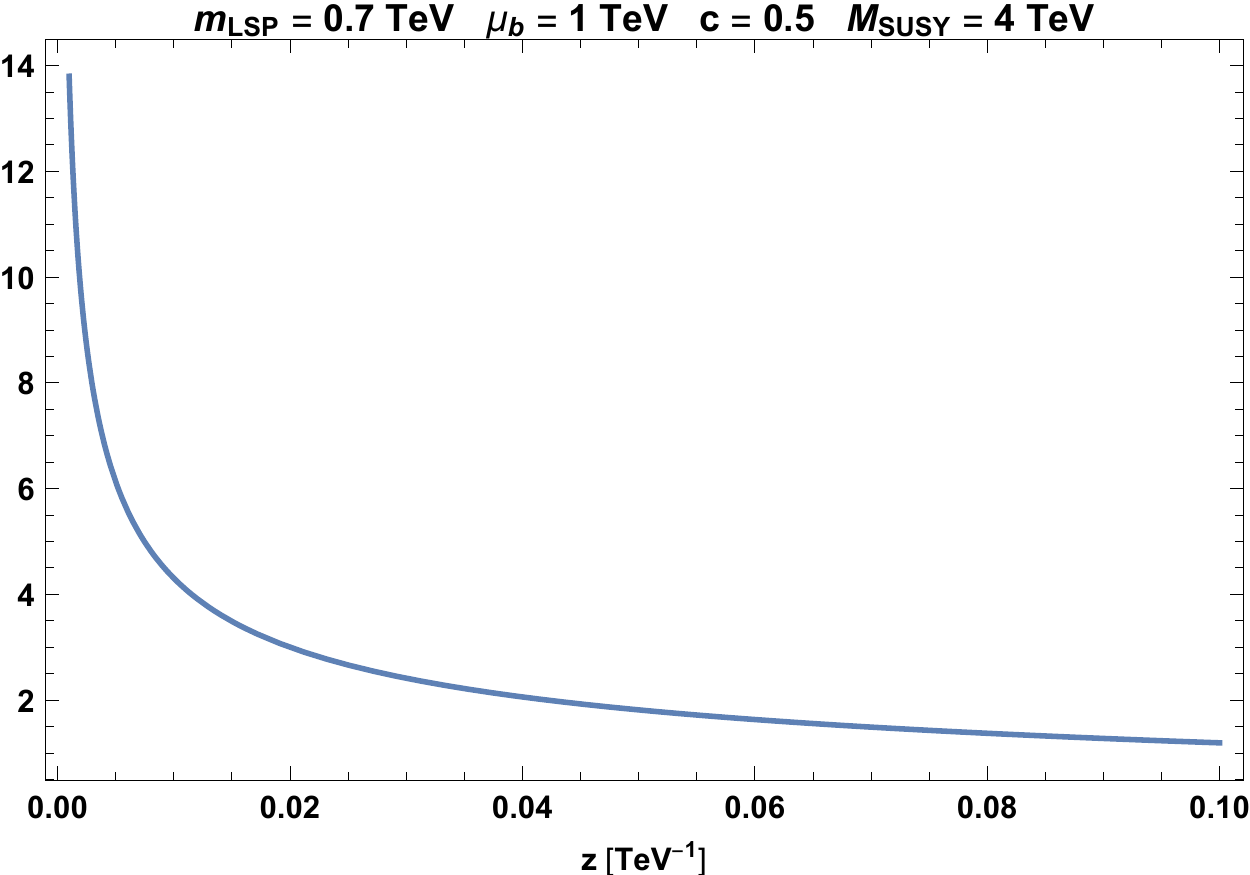}
\caption{A normalized zero mode profile with $M'_{SUSY}\approx 4$ TeV.}
\label{zeromode_wave}
\end{figure}

  Unlike the gluino and squarks, we assume for the sake of simplicity in the simulations that the lifted zero mode of the bino does not merge with its continuum. 
Fig.~\ref{zeromode_wave} shows an example of the zero mode profile with a SUSY breaking mass approximately equal to 4 TeV. The lifted zero mode is clearly peaked on the UV brane.


\subsection{Deconstructing the Continuum}
The spectrum of each field consists of a zero mode and a continuum starting at $\mu$. As discussed in the introduction, we approximate the continuum with a tower of particles. In the literature, this is known as deconstruction \cite{ArkaniHamed:2001ca,Stephanov:2007ry} of the continuum (also see Appendix~\ref{app:KKvsCon}). In the 5D AdS spacetime, we can introduce an IR brane as a regulator and impose Dirichlet and Neumann boundary conditions for the RH and LH fields respectively. Then the 4D effective action is found by substituting the 5D field by a sum over the infinite KK modes. For example, the gluino, denoted by $\tilde{g}$, has a profile like  $\lambda_1$ in \eqref{gaugefields}, i.e.
\beq \label{eq:kkexpansion}
\tilde{g} (x,z)=\sum_n \tilde{g}^n_4(x)e^{\mu_g z}\Big(\frac{z}{z_{UV}}\Big)^{2}\NN_{\tilde{g}_n}(\mu_g,m_{\tilde{g}_n})\mathfrak{h}_L(m_{\tilde{g}_n},z)~.
\eeq
Where $\tilde{g}^n_4(x)$ are the 4D fields
, $\mathfrak{h}_L$ is given by \eqref{solutions3} evaluated at $c=1/2$, and the normalization $\NN_{\tilde{g}_n}$ is fixed by \eqref{normalization}. 
The mass spectrum is found by imposing
\beq
\mathfrak{h}_R(m_{\tilde{g}_n},z_{IR})=0~.
\eeq

The same equation is imposed on every super-multiplet in the 5D theory. The spectrum of each field then consists of a zero mode (which exists even without an IR brane) plus an infinite tower of KK modes with masses starting at the corresponding threshold $\mu_i$, with a spacing roughly given by $\pi/z_{IR}$. For KK modes far from the threshold we can approximate the masses by
\begin{equation}\label{kkmass}
m_n^2\approx \mu^2+\left(\frac14+n\right)^2\pi^2/z_{IR}^2 ~.
\end{equation}

SUSY breaking changes the boundary conditions, and the mass spectrum, however, since $M^2_{SUSY}z_{UV}\ll \mathcal{O}$(1TeV), the KK modes' wave-functions, and hence the continuum spectra, are well approximated by the formulae without SUSY breaking. The main effect of SUSY breaking in these models is to simply lift the zero modes of the superpartners \cite{09terning}.


\section{Production and Decay of KK modes}\label{prodanddecay}
In this section we consider production of the gluino continuum at the LHC and the subsequent decays.
In our toy model we assume, for simplicity, that all leptons live on the UV brane, hence have no continuum modes. The gluon, gluino, quarks, and squarks all have a continuum modes. As explained above, the gluino and squark zero modes are assumed to merge with their continuum. Since the LSP in our model is a zero mode Bino, the electro-weak gauge bosons, the Higgs, and their partners live in the bulk and come with their own continuums. Whether the continuum of these particles significantly change gluino production and decay rates requires a much more detailed analysis. Since we are aiming at a more qualitative analysis, here we will simply ignore them. Alternatively, we can assume all the electroweak zero modes but the LSP's are merged with their continua, and furthermore that the thresholds are large, such that the production of these continua is suppressed.

\begin{figure}[h]
\centering
\begin{fmffile}{gluino1}
   \begin{fmfgraph*}(150,100)
   \fmfleftn{i}{2}
   \fmfrightn{o}{6}
   \fmf{gluon}{i1,v1,i2}
   \fmf{plain,label=$\tilde{g}$,label.side=left}{v2,v1}
    \fmf{plain,label=$\tilde{g}$,label.side=left}{v1,v3}
    \fmf{plain}{o1,v2,o2}
    \fmf{plain}{o5,v3,o6}
    \fmf{plain,fore=red}{o3,v2}
    \fmf{plain,fore=red}{o4,v3}
    \fmfblob{.1w}{v1,v2,v3}
    \fmffreeze
    \fmf{zigzag,width=0.01}{v2,v1}
        \fmf{zigzag,width=0.01}{v1,v3}
\fmf{zigzag,width=.01,fore=red}{v2,o3}
\fmf{zigzag,width=.01,fore=red}{v3,o4}
\fmf{dots}{o1,o2}
\fmf{dots}{o5,o6}
 \end{fmfgraph*}
\end{fmffile}
\caption{Pair production and decay of gluino continuum at the LHC. The red lines correspond to LSP.}
\label{production}
\end{figure}
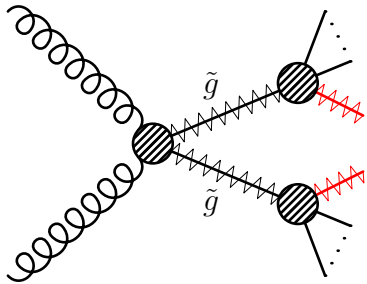

\begin{figure}[H]
\centering
\begin{fmffile}{decaychainexample}
   \begin{fmfgraph*}(300,100)
 \fmfstraight
   \fmfleftn{i}{3}
   \fmfrightn{o}{3}
   \fmftopn{t}{10}
   \fmf{zigzag,width=.01}{i2,v1}
       \fmf{dashes,fore=blue,label=$\tilde{q}_m$}{v1,v2}  
      \fmf{zigzag,width=.01,label=$\tilde{g}_k$}{v2,v3}
   \fmf{dbl_dashes,width=2,label=$\tilde{q}$,label.side=right,fore=blue}{v3,v4}
       \fmf{plain,fore=red}{v4,o2}
   \fmffreeze
      \fmf{zigzag,width=.01,fore=red}{v4,o2}
             \fmf{plain}{v1,i2}
             \fmf{plain}{v2,v3}
       \fmf{plain,label=$q_0$,label.side=left}{v1,t4}
     \fmf{plain,label=$q_0$,label.side=left}{v2,t6}
     \fmf{plain,label=$q_0$,label.side=left}{v3,t8}    
       \fmf{plain,label=$q_0$,label.side=left}{v4,t10}  
        \fmflabel{$\tilde{g}_n$}{i2}
        \fmflabel{$\chi$}{o2}
         \fmfv{decor.shape=circle,decor.filled=full,decor.size=3thick}{v1}
         \fmfv{decor.shape=circle,decor.filled=full,decor.size=3thick}{v2}
          \fmfv{decor.shape=circle,decor.filled=full,decor.size=3thick}{v3}
         \fmfv{decor.shape=circle,decor.filled=full,decor.size=3thick}{v4}
 \end{fmfgraph*}
\end{fmffile}
\caption{Typical KK gluino decay. Double-dash means offshell squarks.}\label{typicalgluinodecay}
\end{figure}
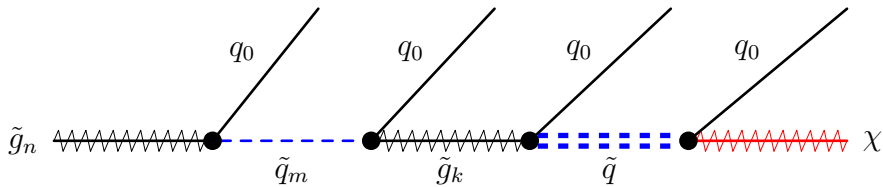

 The process that we are interested in is shown in Fig.~\ref{production}, where $\tilde{g}$ is part of the gluino continuum. The decays of the KK gluinos are calculated in ref.~\cite{11terning}, which we briefly review here. 
We also demonstrate how the decays are \emph{independent} of the choice of the regulator $z_{IR}$ as long as $\mu_g z_{IR}\gg 1$.
The main result of this section is that, using detailed calculations, we show that a typical decay chain for KK gluino is similar to the process depicted in Fig.~\ref{typicalgluinodecay}.
%


\subsection{Production of a Gluino Continuum }\label{sec:prodKK}

\begin{figure}[h]
\centering
\includegraphics[scale=0.7]{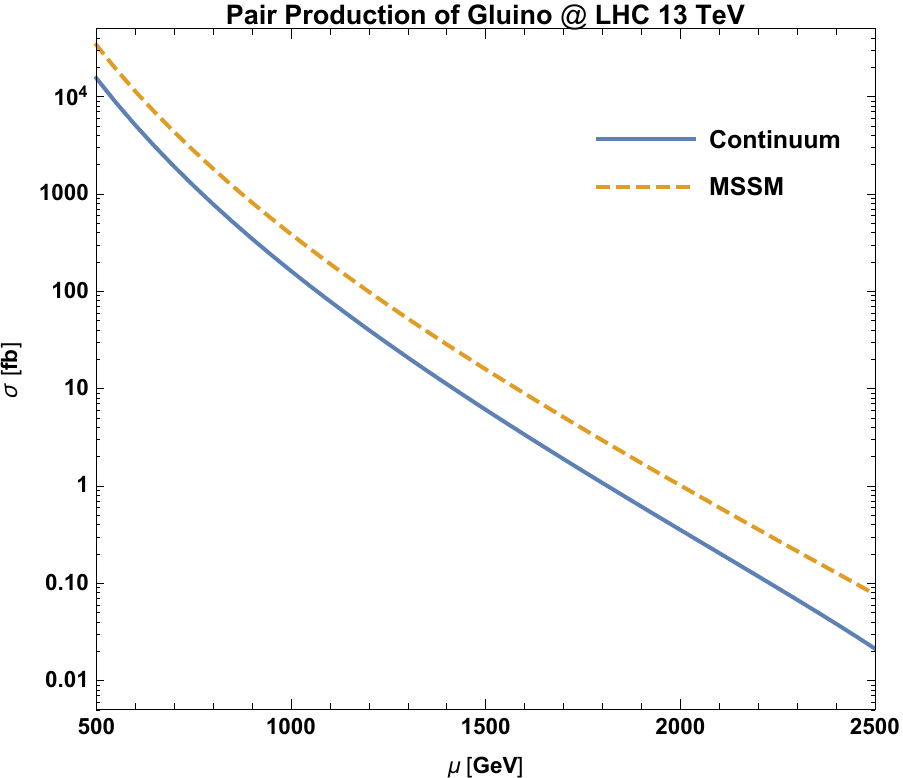}
\caption{Cross section for pair production of a gluino continuum as a function of the threshold $\mu$, assuming $z_{UV}=10^{-3}$ TeV$^{-1}$, $z_{IR}=0.1$ GeV$^{-1}$, $c=0.5$. For comparison, the MSSM gluino pair production cross section (with gluino mass $\mu$)  is also shown.}
\label{xsec13}
\end{figure}

We choose the standard spectral density normalization such that $\int ds\, \rho(s) =1$ \cite{Espinosa:1998xj, 01Akhoury}. We have checked that for our normalization choice \eqref{normalization} this is satisfied. In the KK picture, the gluon's coupling to two KK gluinos are diagonal in KK modes, therefore, the production of one pair of KK gluinos takes the same form as the production of MSSM gluinos of the same mass, which will be denoted as $\sigma_n$. To get the production cross section of a gluino continuum, we need sum $\sigma_n$ carefully. Since the number of KK modes in the sum depends on the IR regulator, we must take into account of the dependence of the mass of the KK gluino on the KK number. Applying chain rule:

\begin{equation}
\sigma =\sum_n\sigma_n=\sum_n\Delta n\frac{\Delta \sigma_n}{\Delta m_n} \frac{\Delta m_n}{\Delta n}
\end{equation}
Taking derivative of \eqref{kkmass} gives us an estimate of $\frac{\Delta m_n}{\Delta n}$. Therefore,
\begin{equation}
\sigma
=\sum_n( \frac{ \pi\Delta n}{z_{IR}} ) \frac{\Delta \sigma_n}{\Delta m_n}\sqrt{1-\frac{\mu^2}{m_n^2}}
\end{equation}
In the continuum limit,
\begin{equation}
\sigma \to \int ^{E_{max}}_{\mu} dm \frac{d \sigma_n}{d m} \sqrt{1-\frac{\mu^2}{m^2}}
\equiv \int ^{E_{max}}_{\mu}dm \frac{d \sigma}{d m} 
\end{equation}
where we took $( \frac{ \pi\Delta n}{z_{IR}} )\approx\Delta m\to dm $.
$E_{max}$ is determined by the partonic center of mass energy reached in a given experiment. The collider cross section must include the convolution with initial parton densities.
 
Fig.~\ref{xsec13} shows the cross section for pair production of gluino continuum as a function of $\mu$ at the LHC with $\sqrt{s}=13$ TeV.


\subsection{2-Body Decay of KK Gluinos}

\begin{figure}
\centering
\includegraphics{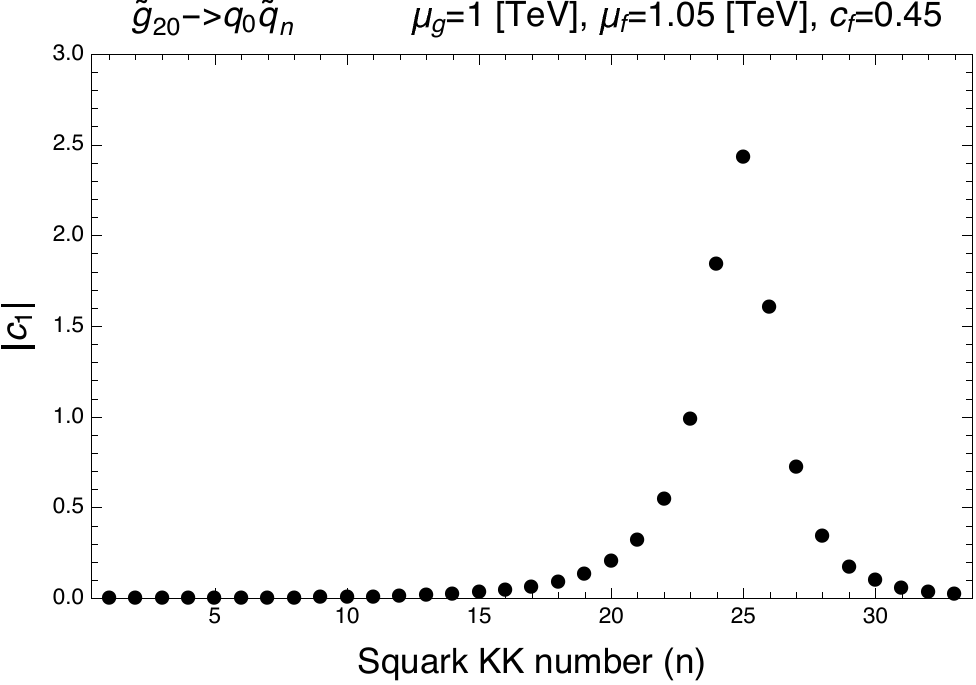}
\caption{The coupling \eqref{eq:gluinosquarkcoupling} between the 20th KK gluino, zero mode quark, and KK squarks, with $z_{IR}=0.1$ GeV$^{-1}$.}
\label{fig:gluinosquarkcoupling}
\end{figure}

A KK gluino couples to KK squarks and KK quarks. The coupling can be found by gauging the matter action and requiring SUSY invariance, and is given by
\beq
\mathcal{L}_{int}&=&\int d^4 x dz \left(\frac Rz \right)^3\int d\theta^4 \Phi^*e^{g_5T^aV^a}\Phi\nonumber \\
&& \supset \int d^4 x dz \left(\frac Rz \right)^3 \frac{ g_5} {\sqrt{2}}(\tilde{g}^{\dagger a}q^{\dagger}T^a\tilde{q} +h.c.)
\eeq
The KK couplings are then found by applying the field decompositions as in \eqref{eq:kkexpansion}. Let us first examine the coupling between a KK gluino, a KK squark, and a SM quark. The relevant term in the Lagrangian is
\begin{equation}
\begin{split}
\mathcal{L}_{int} \supset
&  \frac{ g_5}{\sqrt{2}} T^a \sum_{n,m} c_1(m_{\tilde{g^n}},m_{\tilde{q}^m}) \bar{\tilde{g}}_4(p_m)\tilde{q}_{4}(k_n)\bar{q}^0_4(k_n-p_m)~,
\end{split}
\end{equation}
where the vertex coefficient can be read off as:
\beq\label{eq:gluinosquarkcoupling}
\begin{split}
c_1(m_{\tilde{g}^n},m_{\tilde{q}^m})=&\mathcal{N}^*_{\tilde{g}}(\mu_g,m_{\tilde{g}^n}) \mathcal{N}^*_{q}(\mu_f,0) \mathcal{N}_{\tilde{q}}(\mu_f,m_{\tilde{q}^m}) \times\\
&\int_{z_{UV}}^{z_{IR}}dz \left(\frac{z}{R}\right)^{1/2}e^{(\mu_g-\mu_f)z}\left(\frac{z}{z_{UV}}\right)^{-c}\mathfrak{f}_L(m_{\tilde{q}^m},z)^*\mathfrak{h}_L(m_{\tilde{g}^n},z)~.
\end{split}
\eeq
Here $\mu_g$ is the gluino threshold, and $\mu_f$ is the quark/squark threshold.
For a typical KK gluino, we plot its couplings with different KK squarks and a SM quark in 
Fig.~\ref{fig:gluinosquarkcoupling}.

\begin{figure}
\centering
\begin{subfigure}[b]{0.45\textwidth}
\includegraphics[width=\textwidth]{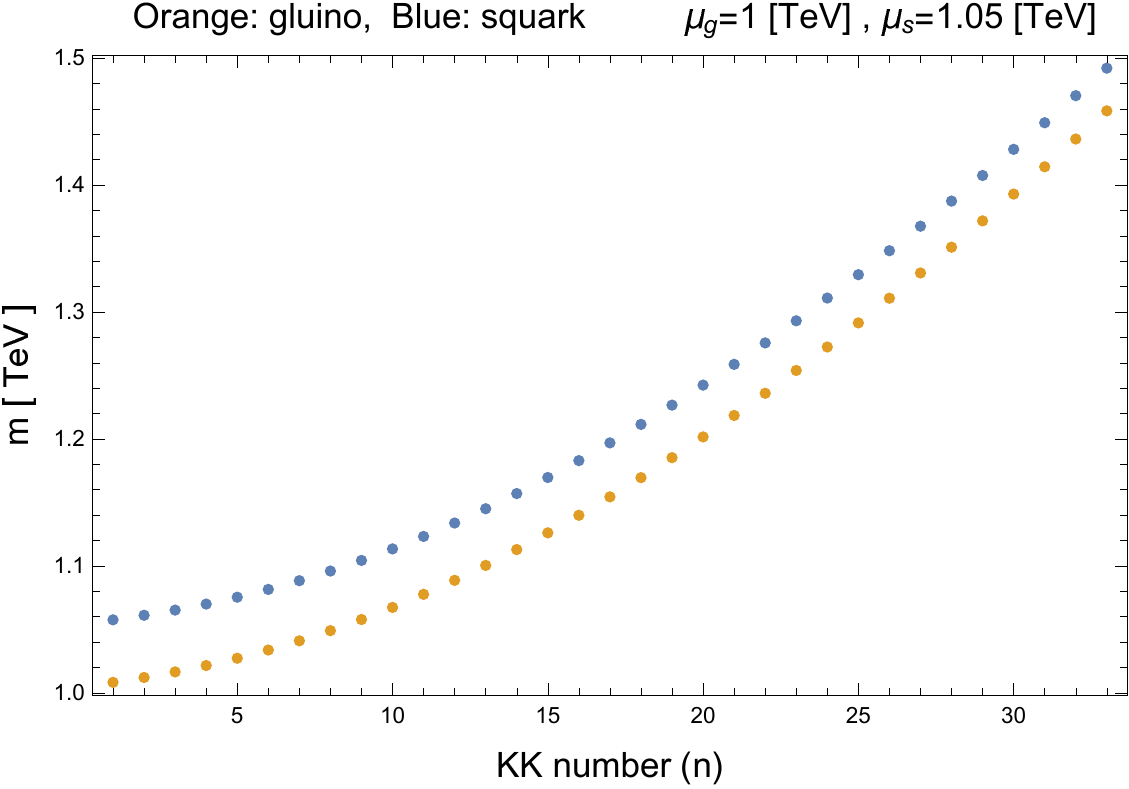}
\subcaption{$z_{IR}=0.1$ GeV$^{-1}$}
\end{subfigure}
\begin{subfigure}[b]{0.45\textwidth}
\includegraphics[width=\textwidth]{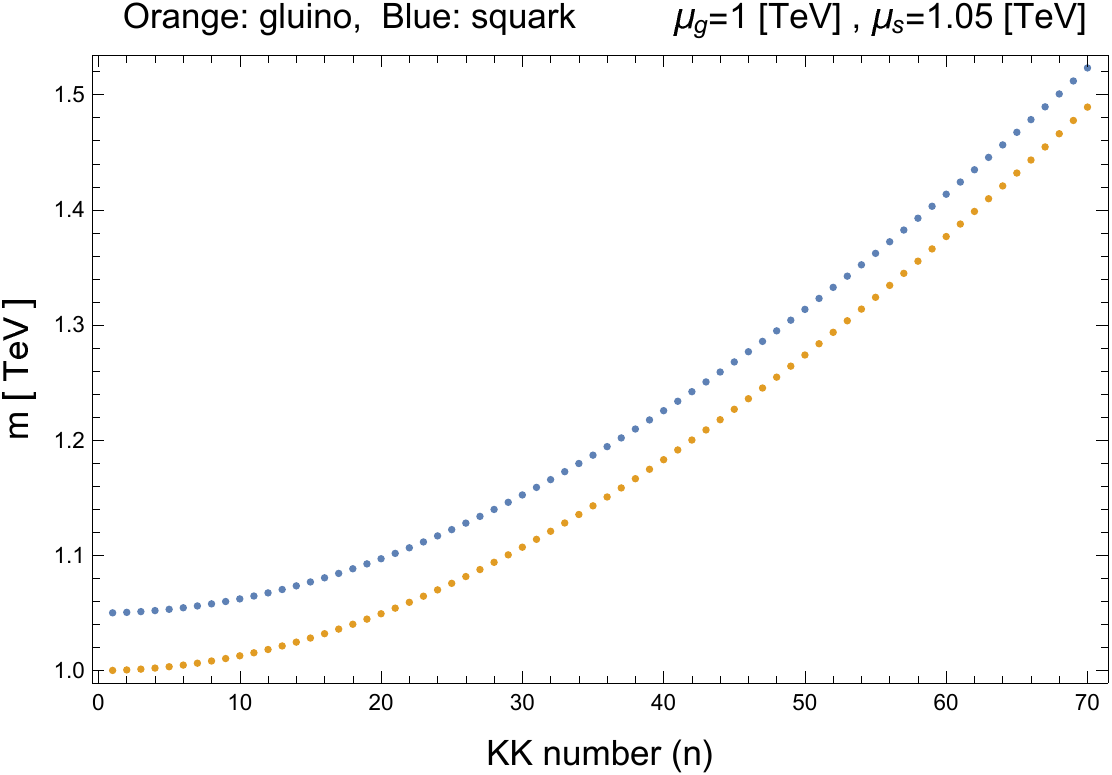}
\subcaption{$z_{IR}=0.2$  GeV$^{-1}$}
\end{subfigure}
\caption{Gluino and squark KK spectrum. 
}
\label{spectrum}
\end{figure}

Because of the factor $\exp(\mu_g-\mu_f)$, the coupling diverges for $\mu_g >\mu_f$. There is nothing wrong with this choice of parameters, this divergence signals that the coupling becomes strong in the IR and hence a simple AdS calculation becomes unreliable. To keep things simple, we stick to the case $\mu_f>\mu_g$ in this paper. Fig.~\ref{spectrum} shows the KK spectra for the gluino and squark for different choices of $z_{IR}$. As we mentioned before, the mass spacing is roughly proportional to $\pi/z_{IR}$.

It was also shown in \cite{11terning} that the biggest contribution to the decay comes when the intermediate particle is produced on-shell.  Therefore, once a highly excited ($n\gg 1$) KK gluino is produced, it undergoes 2-body decay to a lighter KK squark and a SM quark, and the KK squark can further decay to a lighter KK gluino and a SM quark, and the decay continues until the last KK gluino that can no longer decay to any KK squarks on-shell. We address how these lighter states decay in the next subsection.

\begin{figure}[h]
    \centering
    \includegraphics[width=0.9\textwidth]{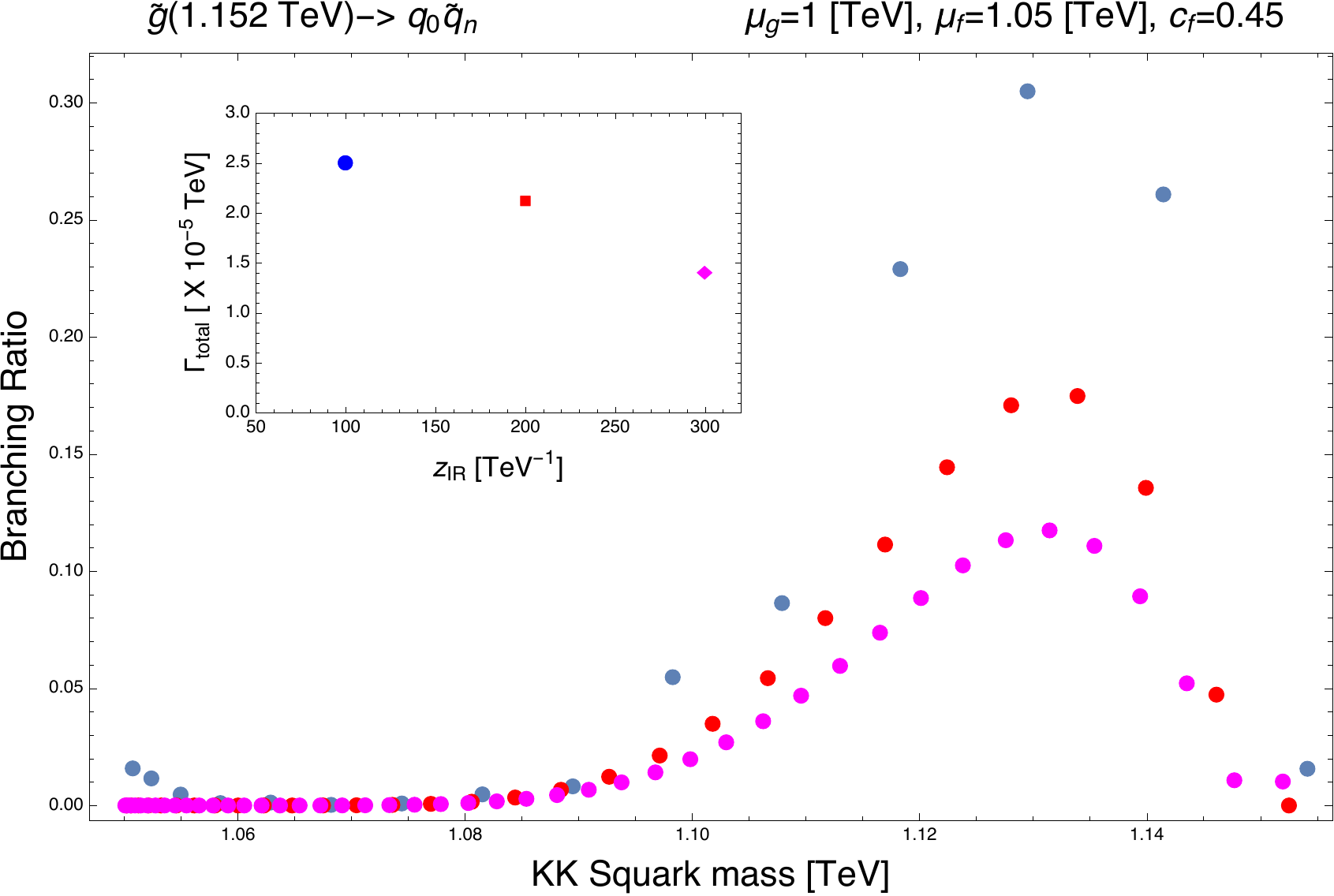}
   \caption{Examples of decay probabilities of one KK gluino with mass $>\mu_f$, decaying to the zero mode quark and a lighter KK squark. The branching ratios for different choices of $z_{IR}$: 0.1 GeV$^{-1}$ (blue), 0.2 GeV$^{-1}$ (red), and 0.3 GeV$^{-1}$ (pink) are shown. The subplot shows the decrease in the total decay width as $z_{IR}$ is increased.}\label{fig:pdfgluino}
\end{figure}

In the rest frame of the gluino, the  partial width for the  $\tilde{g}^m\to q_L^0\tilde{q}_{L}^{* \,n}$ decay is given by
\beq
\Gamma_{\tilde{g}^n\to\tilde{q}^mu}&= \frac{m_{\tilde{g}^n}}{32\pi}\Big(1-\frac{m_{\tilde{q}^m}^2}{m_{\tilde{g}^n}^2}\Big)^2g_5^2\text{Tr}[T^aT^a]|c_1(m_{\tilde{g}^n},m_{\tilde{q}^m})|^2
\eeq
where $T^a$ are the $SU(3)$ generators in the fundamental representation and $\text{Tr}[T^aT^a]=\frac12$. 
For KK gluinos that can decay to KK squarks on-shell, the 2-body decay above will be the dominant contribution to their total widths. Hence, the branching ratio for a given KK gluino decaying to the $m$th squark is approximately
\begin{equation}
\text{BR}(\tilde{g}^n\to\tilde{q}^m)=\frac{\Gamma_{\tilde{g}^n\to\tilde{q}^mu}}{\sum_m \Gamma_{\tilde{g}^n\to\tilde{q}^mu}}
\end{equation}
where the summation is over the kinematically allowed KK squark states.
In Fig.~\ref{fig:pdfgluino}, we show examples of the branching ratios of the same gluino with different choices of $z_{IR}$. As we can see, the mother particle has a higher probability of decaying to the KK mode with a mass close to its own mass, but not to the KK modes with very small mass differences since there is a phase space suppression in the final state. While the partial widths are $z_{IR}$ dependent, the branching fraction for the mother gluino to decay to the KK modes in an \emph{fixed energy interval} remains the same; this is the physical, regulator-independent, quantity we are looking for. In the continuum picture, this translates to the statement that if we produce a KK continuum with $p^2=1.52^2$ TeV$^2$, it will most probably decay to that part of the squark continuum with $p^2\approx 1.13^2$ TeV$^2$.

We also observe that the total decay width of the gluino decreases as $z_{IR}$ increases. This is because that the normalization for each regulated KK mode is proportional to $\Delta=1/z_{IR}$~\cite{Stephanov:2007ry}. As a result, the KK gluino's partial width to a KK squark and a zero mode quark is proportional to $\Delta^2$, whereas the total width of the gluino should be proportional to $\Delta$ since there are roughly $1/\Delta$ KK states that it can decay to. 

\subsection{2-Body Decay of KK Squarks}
\begin{figure}[h]
    \centering
    \includegraphics[width=0.7\textwidth]{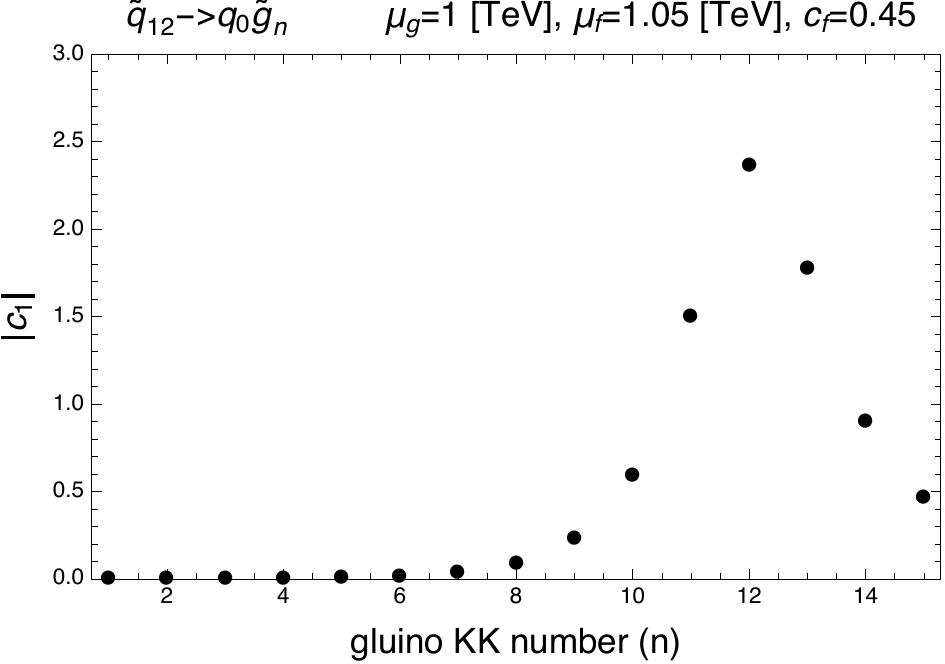}
   \caption{The coupling \eqref{eq:gluinosquarkcoupling} between the 12th KK squark, zero mode quark, and KK gluino, with $z_{IR}=0.1 $GeV$^{-1}$.}\label{fig:squarkgluinocoupling}
\end{figure}

Turning our attention to KK squarks, we find a similar behavior in their decays to KK gluinos. The coupling is given by \eqref{eq:gluinosquarkcoupling} and
is plotted as a function of the KK number for the gluino in Fig.~\ref{fig:squarkgluinocoupling}. 
The decay width is given by
\beq
\Gamma_{\tilde{q}^m\to \tilde{g}^nu}&= \frac{m_{\tilde{q}^n}}{32\pi}\Big(1-\frac{m_{\tilde{g}^m}^2}{m_{\tilde{q}^n}^2}\Big)^2g_5^2\text{Tr}[T^aT^a]|c_1(m_{\tilde{g}^m},m_{\tilde{q}^n})|^2~.
\eeq

Furthermore, the squark can decay to a quark together with other electroweak-inos, namely bino, zino, and wino. For simplicity, we will assume that the threshold for the electroweak sector is of order of few TeV, though it is not necessarily excluded at lower values.  
We further assume that the zero modes for the superpartners are all merged into their continua except the bino, which will play the role of the LSP in our model. 
The coupling of the KK squark to a bino and a SM quark is given by
\beq\label{int:bino}
\LL_{int} =\int d^4xdz \sqrt{g}g'_5[\tilde{B}^{\dagger}q^{\dagger}Y\tilde{q}+h.c.].
\eeq
$g'_5$ can be related to 4D electroweak coupling by noting that 5D photon couples to leptons that are localized on the UV brane by
\beq
\LL_{int} =\int d^4xdz \sqrt{g}g'_5\left(\frac{z}{R}\right)\delta(z-z_{UV}) \bar{l}\gamma^{\mu}A_{\mu} l~.
\eeq
After substituting the zero modes and doing the integral we find: 
\beq
{g'_5}^2=g_{w}^2 R[\log \frac{1}{2R\mu_b}-\gamma_E]~.
\eeq
The bino zero mode takes a similar form to the zero mode gluino
\begin{equation}\label{binoprofile}
\tilde{B}^0(p,z)=\tilde{\chi}e^{\mu_b z} \left(\frac zR\right)^{2} h^0_L(m_{\tilde{\chi}},z)~,
\end{equation} 
where $h_L^0$ is given by \eqref{bino}. The 4D field $\tilde{\chi}$ is the LSP.  

\begin{figure}[H]
    \centering
    \includegraphics[width=0.7\textwidth]{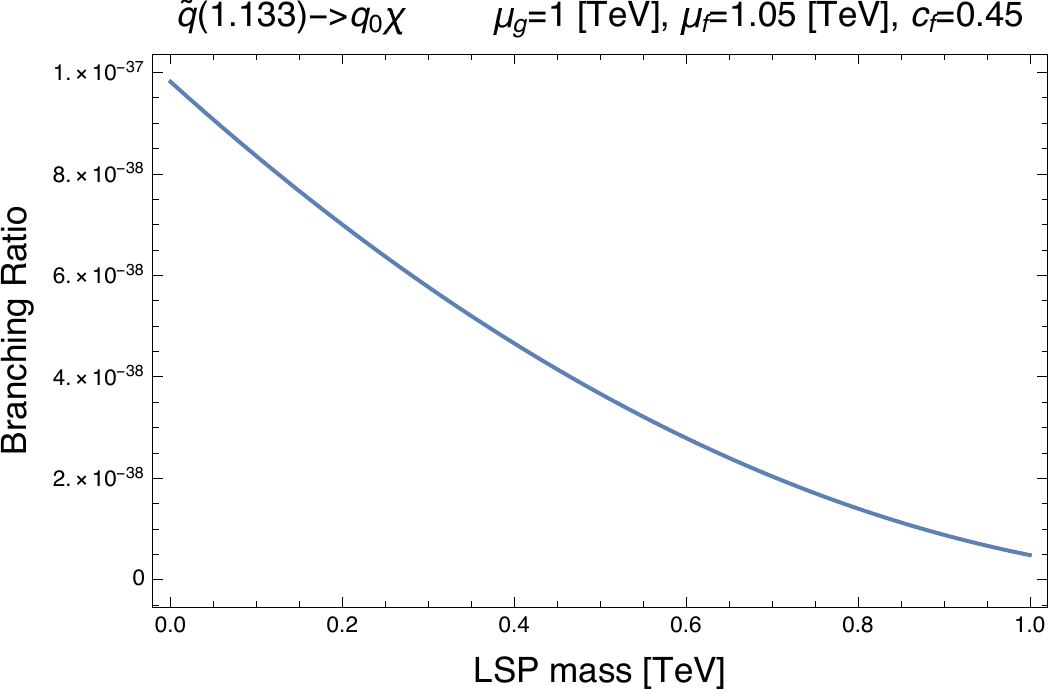}
   \caption{Branching ratio of KK squark decaying to a SM quark and LSP as a function of LSP mass.   }\label{fig:pdfsquarktolsp}
\end{figure}
\begin{figure}[h]
    \centering
    \includegraphics[width=0.7\textwidth]{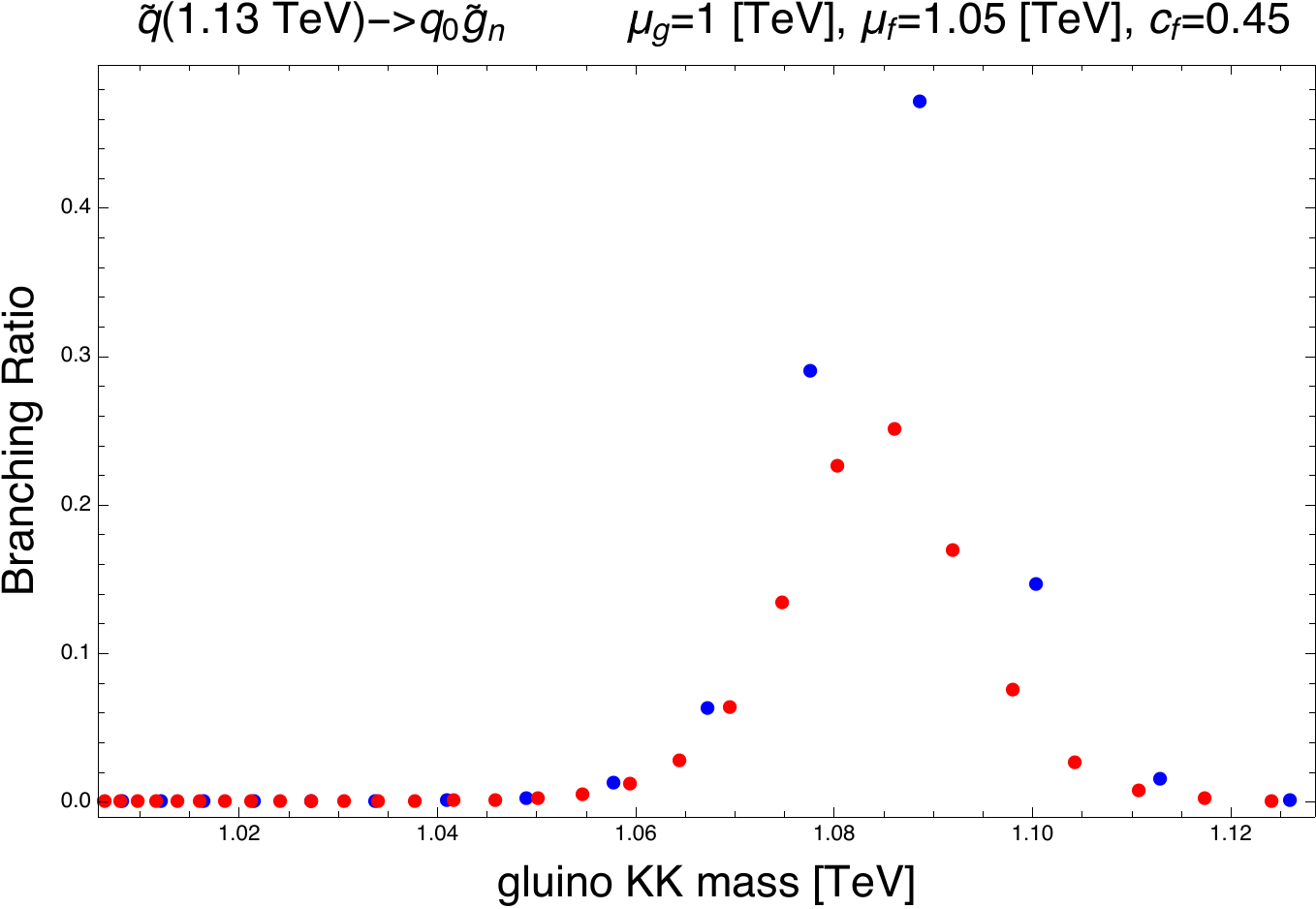}
   \caption{Examples of decay probabilities of a KK squark with mass $>\mu_g$, decaying to the zero mode quark and a lighter KK gluino. The BR is plotted for different choices of $z_{IR}$: 0.1 GeV$^{-1}$ (blue), and 0.2 GeV$^{-1}$ (red).}\label{fig:pdfsquark}
\end{figure}

After substituting the bino zero mode and quark zero mode and squark KK mode profiles into \eqref{int:bino}, we find
\beq
\LL_{int}=g'_5 Y \sum_n c_2(m_{\tilde{q}_n},m_{\chi})\tilde{q}_n(p)\bar{\chi}(k)q^0(k-p)~,
\eeq
with 
\begin{align}
c_2(m_{\tilde{q}_n},m_{\chi})&=\NN_{\tilde{q}}(\mu_f,m_{\tilde{q}_n})\NN_q(\mu_f,0)\NN_{\chi}(m_{\chi})\times \nonumber\\
\int dz
& (\frac{R}{z})^5[\left(\frac{z}{R}\right)^{3/2}\mathfrak{f}_L(m_{\tilde{q}_n},z)][\left(\frac{z}{R}\right)^2e^{-\mu_f z}(\frac{z}{z_{UV}})^{-c}][e^{\mu_b z}\left(\frac{z}{R}\right)^2\mathfrak{h}_L(m_{\chi},z)]\nonumber\\
&=\NN_{\tilde{q}}(\mu_f,m_{\tilde{q}_n})\NN_q(\mu_f,0)\NN_{\chi}(m_{\chi})\times\nonumber\\
& \quad \int dz \left(\frac{z}{R}\right)^{1/2}e^{(\mu_b-\mu_f )z}\left(\frac{z}{z_{UV}}\right)^{-c}\mathfrak{f}_L(m_{\tilde{q}_n},z)\mathfrak{h}_L(m_{\chi},z)~,
\label{csqgx}
\end{align}
where we approximated the zero mode profile by
\beq
\mathfrak{h}_L(m_{\chi},z)\approx \left(\frac{z}{z_{UV}}\right)^{-1/2}e^{-\mu_b z}~.
\eeq
The zero mode bino normalization is the same as that of photon, so we have the relation $g'_5 \NN_{\chi}=g_{w}$. Due to the fact that the zero mode bino is localized close to the UV brane, its coupling to KK squarks is very small, thus the squark's decay to the LSP is negligible, as shown in Fig.~\ref{fig:pdfsquarktolsp}.
In Fig.~\ref{fig:pdfsquark} we show the decay probability of a given KK squark to KK gluinos with smaller masses.


\subsection{3-Body Decay of a Light KK Gluino}
The KK gluinos that are lighter than the lightest KK squark can no longer decay to on-shell KK squarks. Instead, they decay to two SM quarks and a bino (LSP) or a lighter KK gluino via off-shell KK squarks. For calculational purposes, it is easier to use the 5D holographic propagator for the squark instead of summing over infinite number of off-shell KK squarks. 
The 5D squark propagator \cite{Falkowski:2008yr} (the Neumann propagator) is given by
\begin{equation}
iP(p,z,z')=\theta(z'-z)[\Sigma_{F}K(p,z)K(p,z')-S(p,z)K(p,z')]+(z\leftrightarrow z')~,
\end{equation}
where $\Sigma_F$ is the kinetic term 
calculated in Appendix \ref{app:holoAction}.
The holographic profile, $K$ is given by
\begin{equation}
K(p,z)=\Big(\frac{z}{z_{UV}}\Big)^{\frac 32}\frac{W\Big(\kappa,\frac12+c,2\sqrt{\mu_f^2-p^2}z\Big)}{W\Big(\kappa,\frac12+c,2\sqrt{\mu_f^2-p^2}z_{UV}\Big)}~.
\end{equation}
The other function, $S$, is
\begin{equation}
\begin{split}
S(p,z)=&\Big(\frac{z}{z_{UV}}\Big)^{\frac 32}
\frac1{2\sqrt{\mu^2_f-p^2}}\frac{\Gamma(1+c-\kappa)}{\Gamma(2+2c)}
\\
&\Bigg[
M\Big(\kappa,\frac12+c,2\sqrt{\mu_f^2-p^2}z\Big)W\Big(\kappa,\frac12+c,2\sqrt{\mu_f^2-p^2}z_{UV}\Big)\\
&-
W\Big(\kappa,\frac12+c,2\sqrt{\mu_f^2-p^2}z\Big)M\Big(\kappa,\frac12+c,2\sqrt{\mu_f^2-p^2}z_{UV}\Big)~\Bigg].
\end{split}
\end{equation}
We will first look at the decay of gluinos to lighter gluinos and next to the LSP.


\subsubsection{$\tilde{g}^n\to q_L^0 \bar{q}_L^0\tilde{g}^m$ }\label{gluinogluinojj}

\begin{figure}[ht!]
\centering

\raisebox{7mm}{\begin{fmffile}{gluino3bodydecaytogluino}
   \begin{fmfgraph*}(140,50)
   \fmfleftn{i}{5}
   \fmfrightn{o}{5}
   \fmf{zigzag,width=.01}{i2,v1}
    \fmflabel{$\tilde{g}_n$}{i2}
   \fmf{dbl_dashes,width=2,label=$\tilde{q}$,label.side=left,fore=blue}{v2,v1}
    \fmf{plain}{o2,v2}
            \fmflabel{$z$}{v1}
         \fmfv{decor.shape=circle,decor.filled=full,decor.size=3thick,label=$z$,label.angle=-70}{v1}
         \fmfv{decor.shape=circle,decor.filled=full,decor.size=3thick,label=$z'$,label.angle=-130}{v2}
    \fmffreeze
     \fmf{plain}{i2,v1}
       \fmf{zigzag,width=.01}{v2,o2}
        \fmflabel{$\tilde{g}_m$}{o2}
        \fmf{plain}{ov,v2}
         \fmf{phantom}{ov,o3}
         \fmffreeze
         \fmfshift{16up}{ov}
    \fmf{plain}{ol1,v1}
        \fmf{phantom}{ol2,ol1}
            \fmf{phantom}{o4,ol2}
    \fmffreeze
         \fmfshift{16up}{ol1}
          \fmflabel{$q_0$}{ov}
        \fmflabel{$q_0$}{ol1}
 \end{fmfgraph*}
\end{fmffile}}
\hspace{12mm}\raisebox{9mm}{$\to$}\hspace{12mm}
\raisebox{3mm}{\begin{fmffile}{gluino3bodydecaytogluinoeffective}
\begin{fmfgraph*}(80,40)
  \fmfleft{i1,i2,i3}
  \fmfright{o1,o2,o3}
  \fmf{plain}{i2,v1}
   \fmflabel{$\tilde{g}_n$}{i2}
  \fmf{plain,tension=0.33}{v1,o1}
    \fmf{plain,tension=0.33}{v1,o2}
  \fmf{plain,tension=0.33}{v1,o3}
   \fmflabel{$\tilde{g}_m$}{o1}
\fmffreeze
  \fmf{zigzag,width=.01}{i2,v1}
  \fmf{zigzag,width=.01}{v1,o1}
      \fmfblob{0.1w}{v1}
      \fmfv{decor.shape=circle,decor.filled=shaded,decor.size=6thick,label=$v_1$,label.angle=+120}{v1}
                \fmflabel{$q_0$}{o2}
        \fmflabel{$q_0$}{o3}
  \end{fmfgraph*}
\end{fmffile}}
\caption{$\tilde{g}^n\to q_L^0 \bar{q}_L^0\tilde{g}^m$ effective vertex (right) from the propagation of squark in 5D (left)}
\label{fig:gluinoqqgluino}
\end{figure}
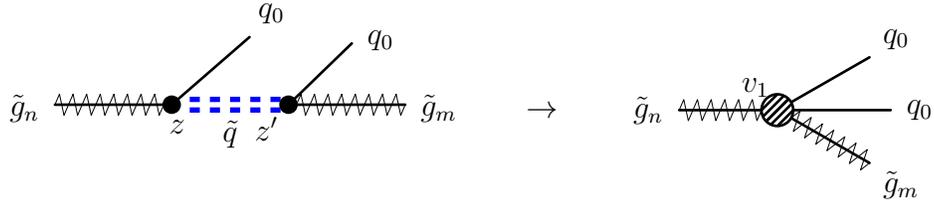

The decay $\tilde{g}^n\to q_L^0 \bar{q}_L^0\tilde{g}^m$ via off-shell KK squarks has been worked out in \cite{11terning}.
The diagram representing this process is shown in fig.~\ref{fig:gluinoqqgluino}.

Integrating over the fifth dimension we find an effective interaction in the 4D theory between two different KK gluinos and two zero mode quarks, 
\begin{equation}
\begin{split}
v_1(m_{\tilde{g}^n},q,m_{\tilde{g}^m})=&\mathcal{N}_{q}^2(\mu_f,0)\mathcal{N}_{\tilde{g}}(\mu_g,m_{\tilde{g}^n})\mathcal{N}_{\tilde{g}}(\mu_g,m_{\tilde{g}^m})\\
&\int _{z_{UV}}^{z_{IR}} dz\left(\frac{z}{z_{UV}}\right)^{-1-c}e^{(\mu_g-\mu_f)z}\mathfrak{h}_L(m_{\tilde{g}^n},z)\\
&\times \int _{z_{UV}}^{z_{IR}}dz'{\left(\frac{z'}{z_{UV}}\right)}^{-1-c}e^{(\mu_g-\mu_f)z'}\mathfrak{h}_L(m_{\tilde{g}^m},z')
P(q,z,z')~,
\end{split} 
\end{equation}
where $q$ is the momentum carried by the squark, which is equal to the sum of the momentum of the $m$th gluino and the second outgoing quark. Note that the mass dimensions are: $[g_5]=-1/2$, $[\NN]=1/2$, $[\mathfrak{h}_L]=0$, and $[P(q,z,z')]=-1$; so the dimension of the $[g_5^2v_1]=-2$, which is the dimension of a propagator. 

The decay rate is given by
\begin{equation}
\begin{split}
\frac{d\Gamma_3(\tilde{g}^n\to q_L^0 \bar{q}_L^0\tilde{g}^m)}{dE_u}=&g_5^4 |v_1(m_{\tilde{g}^n},q,m_{\tilde{g}^m})|^2\times\\
&
\frac{E_u^2(2E_u m_{\tilde{g}^n}-m_{\tilde{g}^n}^2+m_{\tilde{g}^n}^2)^2}{32m_{\tilde{g}^n}(m_{\tilde{g}^n}-2E_u)\pi^3}\frac18\text{Tr}[T^aT^bT^bT^a]
\end{split}
\end{equation}
where Tr$[T^aT^bT^bT^a]=C_AC_F^2=16/3$ for $SU(3)$.

\subsubsection{$\tilde{g}^n\to q_L^0 \bar{q}_L^0\tilde{\chi}^0$}
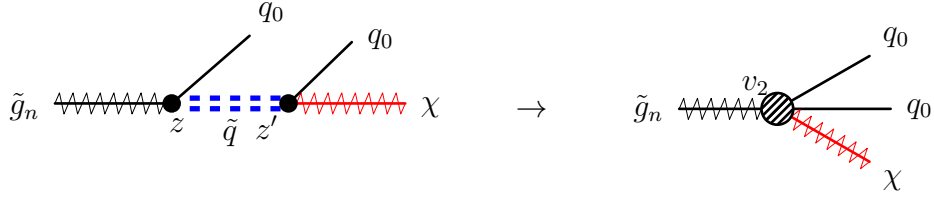
\begin{figure}[ht!]
\centering
\raisebox{7mm}{\begin{fmffile}{gluino3bodydecaytolsp}
   \begin{fmfgraph*}(140,50)
   \fmfleftn{i}{5}
   \fmfrightn{o}{5}
   \fmf{zigzag,width=.01}{i2,v1}
    \fmflabel{$\tilde{g}_n$}{i2}
   \fmf{dbl_dashes,width=2,label=$\tilde{q}$,label.side=left,fore=blue}{v2,v1}
    \fmf{plain,fore=red}{o2,v2}
         \fmfv{decor.shape=circle,decor.filled=full,decor.size=3thick,label=$z$,label.angle=-70}{v1}
         \fmfv{decor.shape=circle,decor.filled=full,decor.size=3thick,label=$z'$,label.angle=-130}{v2}
    \fmffreeze
     \fmf{plain}{i2,v1}
       \fmf{zigzag,width=.01,fore=red}{v2,o2}
        \fmflabel{$\chi$}{o2}
        \fmf{plain}{ov,v2}
         \fmf{phantom}{ov,o3}
         \fmffreeze
         \fmfshift{16up}{ov}
    \fmf{plain}{ol1,v1}
        \fmf{phantom}{ol2,ol1}
            \fmf{phantom}{o4,ol2}
    \fmffreeze
         \fmfshift{16up}{ol1}
        \fmflabel{$q_0$}{ov}
        \fmflabel{$q_0$}{ol1}
 \end{fmfgraph*}
\end{fmffile}}\hspace{12mm}\raisebox{9mm}{$\to$}\hspace{12mm}
\raisebox{3mm}{
\begin{fmffile}{gluino3bodydecaytolspeffective}
\begin{fmfgraph*}(80,40)
  \fmfleft{i1,i2,i3}
  \fmfright{o1,o2,o3}
  \fmf{plain}{i2,v1}
   \fmflabel{$\tilde{g}_n$}{i2}
  \fmf{plain,tension=0.33,fore=red}{v1,o1}
    \fmf{plain,tension=0.33}{v1,o2}
  \fmf{plain,tension=0.33}{v1,o3}
   \fmflabel{$\chi$}{o1}
\fmffreeze
  \fmf{zigzag,width=.01}{i2,v1}
  \fmf{zigzag,width=.01,fore=red}{v1,o1}
      \fmfv{decor.shape=circle,decor.filled=shaded,decor.size=6thick,label=$v_2$,label.angle=+120}{v1}
              \fmflabel{$q_0$}{o2}
        \fmflabel{$q_0$}{o3}
  \end{fmfgraph*}
\end{fmffile}}
\caption{$\tilde{g}^n\to q_L^0 \bar{q}_L^0\tilde{\chi}^0$ effective vertex (right) from the propagation of squark in 5D (left)}
\label{fig:gluinoqqLSP}
\end{figure}

To estimate the 3-body decay width involving the LSP, we assume that the LSP is mainly a zero mode bino $\tilde{B}^0$. The effective 4D vertex is obtained by propagating the final and initial particles into the bulk and connecting the squark-quark-bino vertex to squark-quark-gluino vertex by squark 5D propagator. The diagram representing this process is shown in Fig.~\ref{fig:gluinoqqLSP}.
The calculation of the differential decay width is similar to that of section \ref{gluinogluinojj}.
%
%
%
In the rest frame of the gluino,
\begin{equation}
\begin{split}
\frac{d\Gamma_3(\tilde{g}^n\to q_L^0 \bar{q}_L^0\tilde{\chi}^0)}{dE_u}=&g_5^2 g_5^{\prime 2} |v_2(m_{\tilde{g}^n},q,m_{\tilde{\chi}})|^2\times\\
&
\frac{E_u^2(2E_u m_{\tilde{g}^n}-m_{\tilde{g}^n}^2+m_{\tilde{\chi}}^2)^2}{32m_{\tilde{g}^n}(m_{\tilde{g}^n}-2E_u)\pi^3}\frac18\text{Tr}[T^aT^a]Y^2~,
\end{split}
\end{equation}
where $E_u$ is the energy of the first outgoing quark, and 
\begin{equation}
\begin{split}
v_2(m_{\tilde{g}^n},q,m_{\tilde{\chi}})=&\mathcal{N}_{q}^2(\mu_f,0)\mathcal{N}_{\tilde{g}}(\mu_g,m_{\tilde{g}^n})\mathcal{N}_{\tilde{\chi}}(\mu_b,m_{\tilde{\chi}})\\
&\int _{z_{UV}}^{z_{IR}} dz\left(\frac{z}{z_{UV}}\right)^{-1-c}e^{(\mu_g-\mu_f)z}\mathfrak{h}_L(m_{\tilde{g}^n},z)\\
&\times \int _{z_{UV}}^{z_{IR}}dz'{\left(\frac{z'}{z_{UV}}\right)}^{-1-c}e^{(\mu_b-\mu_f)z'}\mathfrak{h}_L(m_{\tilde{\chi}},z')
P(q,z,z')~.
\end{split}
\end{equation}
Note that the mass dimension  $[g_5g'_5v_2]=-2$, as it should be.

\begin{figure}[h]
\centering
\includegraphics[scale=0.8]{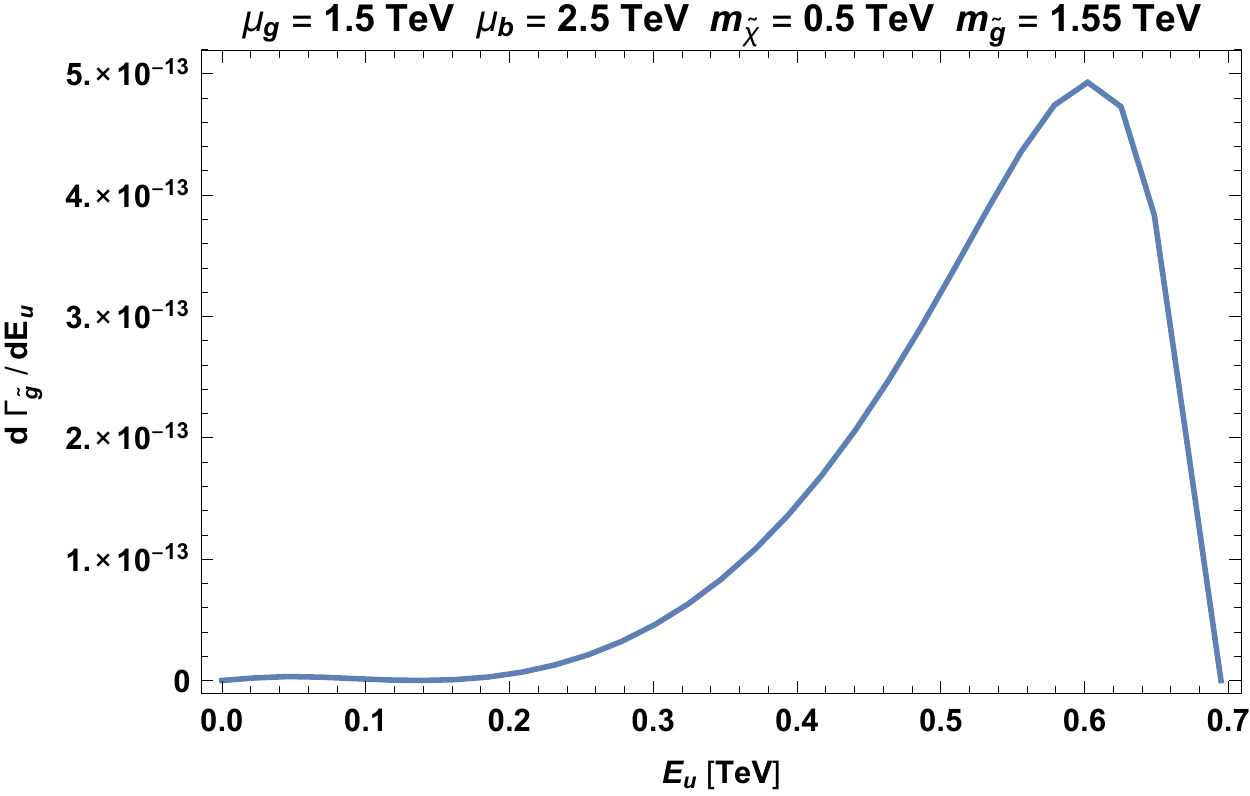}
\caption{Three body differential decay width for the KK gluino of $1.55$ TeV as a function of the energy of the first outgoing quark.}
\label{3body}
\end{figure}

\begin{figure}[H]
    \centering
    \includegraphics[width=1\textwidth]{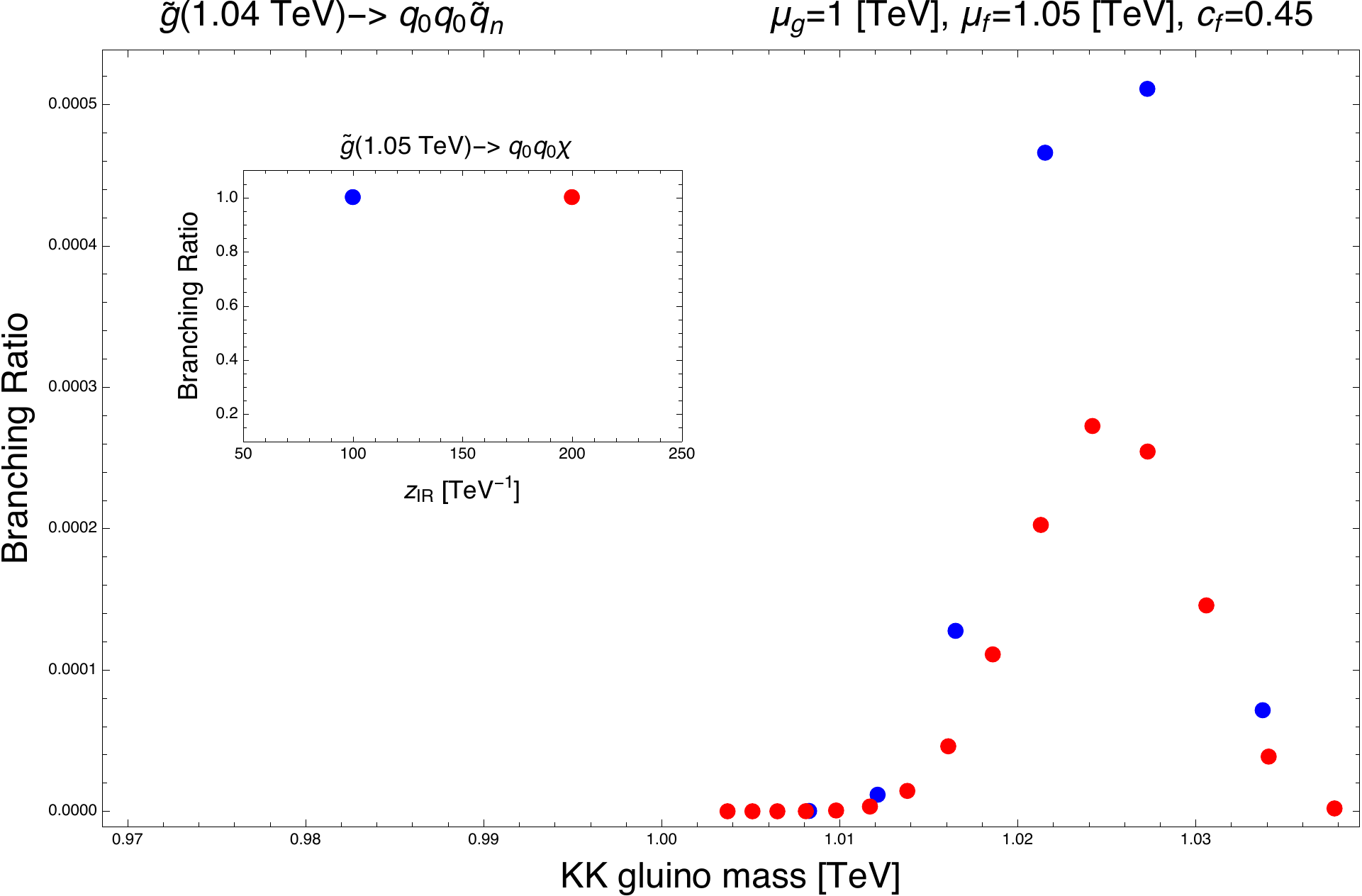}
   \caption{Examples of decay probabilities of a KK gluino with mass $<\mu_f$ for different choices of $z_{IR}$: 0.1 GeV$^{-1}$ (blue), and 0.2 GeV$^{-1}$ (red). The small graph shows the probability of its decay to LSP, which is assumed to be massless.}\label{fig:3bodydecayofgluino}
\end{figure}

Fig.~\ref{3body} shows an example of the differential decay width of a KK gluino of $1.55$ TeV mass, decaying to a neutralino. 
Integrating this differential decay allows us to find the partial width of a KK gluino's decay to the LSP. 
We want to investigate how the branching ratio of the gluino decaying to less massive gluinos compares to its decay to a neutralino. Fig.~\ref{fig:3bodydecayofgluino} shows that these light gluinos predominately decay to LSP.

Let us finish this section by making a comment on the total decay width and decay length of the unparticle. As observed previously (see fig.~\ref{fig:pdfgluino} for example), the decay width of each KK particle is regulator dependent, and falls as a function of $1/z_{IR}$. As we approach the continuum by taking $z_{IR}\to\infty$, each KK mode's decay width vanishes (and the decay length diverges). As a result, one might suspect that the continuum unparticles will not decay at all. This is in fact incorrect. 
In the continuum limit, the individual KK modes are not observable objects. Similar to cancelation of IR divergences in QED or QCD, and similar to our argument regarding the production of KK modes in section~\ref{sec:prodKK}, we need to sum the KK particles over some energy range to find the decay width of the continuum unparticle. 
We can either sum over all the energies bigger than the threshold, $\mu$, to find the total decay width of the unparticle, or, if the unparticle is produced at a given energy, sum over an energy range dictated by the experiment. In both cases, as we remove the IR regulator by taking $z_{IR}\to\infty$, the number of KK modes in an energy interval increases as the decay width of each of them decreases, rendering a finite result for the decay width of the unparticle.

\section{Gluino Continuum at the LHC }\label{lhc}
In this section, we initiate a collider study for gluino continuum decays. We want to investigate whether the gluino continuum reveals itself differently compared to the MSSM gluino at the LHC. Specifically, we focus on a benchmark with a gluino continuum threshold of $\mu_g=1.5$ TeV. The NLO cross section of pair producing the gluino continuum is estimated to be 6 fb at the LHC  with 13 TeV center of mass energy, which is comparable to the pair production cross section of a MSSM gluino with a mass of 1.65 TeV. 
\begin{figure}[h]
    \centering
    \begin{subfigure}[b]{0.45\textwidth}
        \includegraphics[scale=0.7]{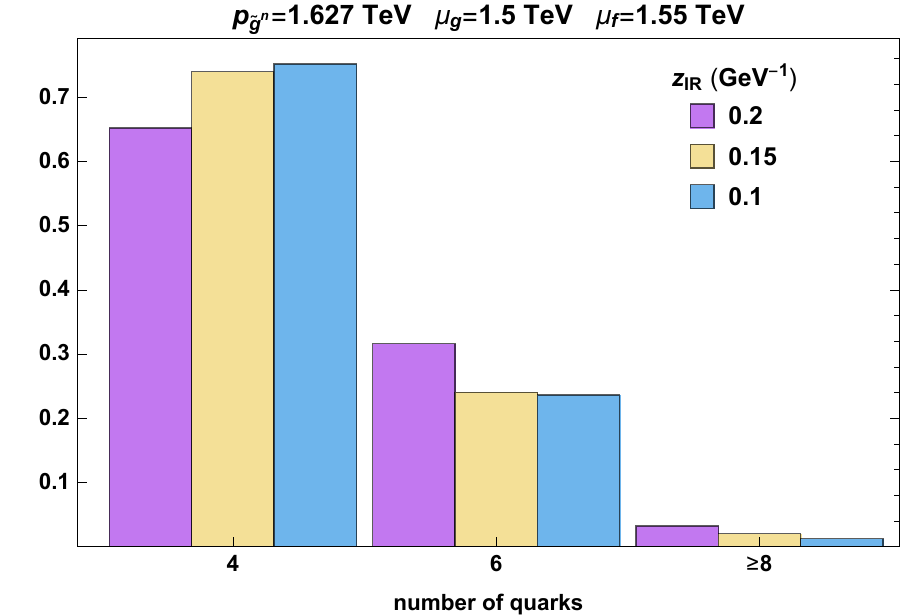}
        \caption{}
        \label{fig:njonekk}
    \end{subfigure}
    ~ 
    \begin{subfigure}[b]{0.45\textwidth}
        \includegraphics[scale=0.7]{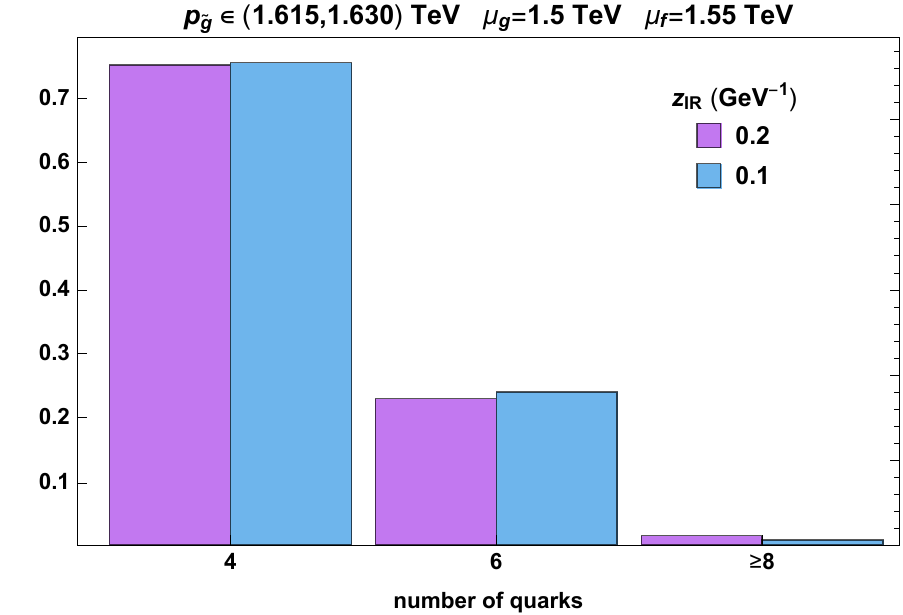}
        \caption{}
        \label{fig:njoneinterval}
    \end{subfigure}
    ~ 
   \caption{Normalized distributions of number of light quarks emitted in the decay of a pair of KK gluinos (a) and KK gluinos in an energy window (b) for different choices of $z_{IR}$. On the right panel the number of KK modes contained in the chosen energy window is one (three) for $z_{IR}=0.1 $GeV$^{-1}$ (0.2 GeV$^{-1}$).}\label{fig:njvszir}
\end{figure}

In order to simulate the signal, we need to pick a regulator $z_{IR}$ so that the gluino continuum can be approximated by a finite KK tower with its upper limit set by the energy reach of the LHC. Fig.~\ref{fig:njvszir} compares the number of light quarks emitted in the decay of a pair of individual KK gluinos for different choices of $z_{IR}$.
The decay is performed step by step according to the KK gluino's and subsequent squark's branching ratios, as explained in Section~\ref{prodanddecay}.
As shown in Fig.~\ref{fig:njoneinterval}, the distribution is almost identical if we sum the KK modes contained in an energy interval instead of just considering a single pair of them (\ref{fig:njonekk}). This observation is in accord with the discussion at the end of Section~\ref{prodanddecay} and led us to using $z_{IR} = 0.1$ GeV$^{-1}$ for our simulations which should provide a very good approximation to the continuum result.

\begin{figure}[h]
    \centering
    \begin{subfigure}[b]{0.5\textwidth}
        \includegraphics[scale=0.54]{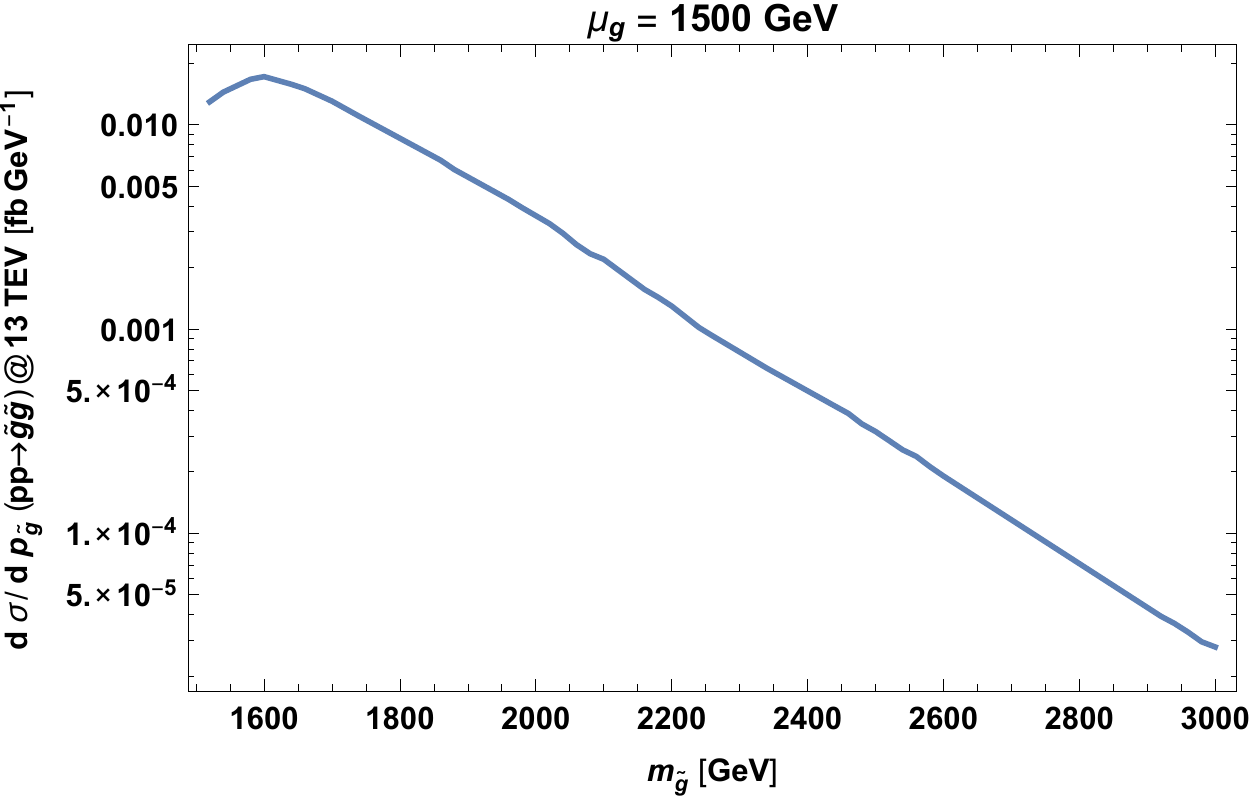}
        \caption{$d\sigma /dm$ vs $p_{\tilde{g}}$.}
        \label{fig:dxsec}
    \end{subfigure}
    ~ 
    \begin{subfigure}[b]{0.45\textwidth}
        \includegraphics[scale=0.5]{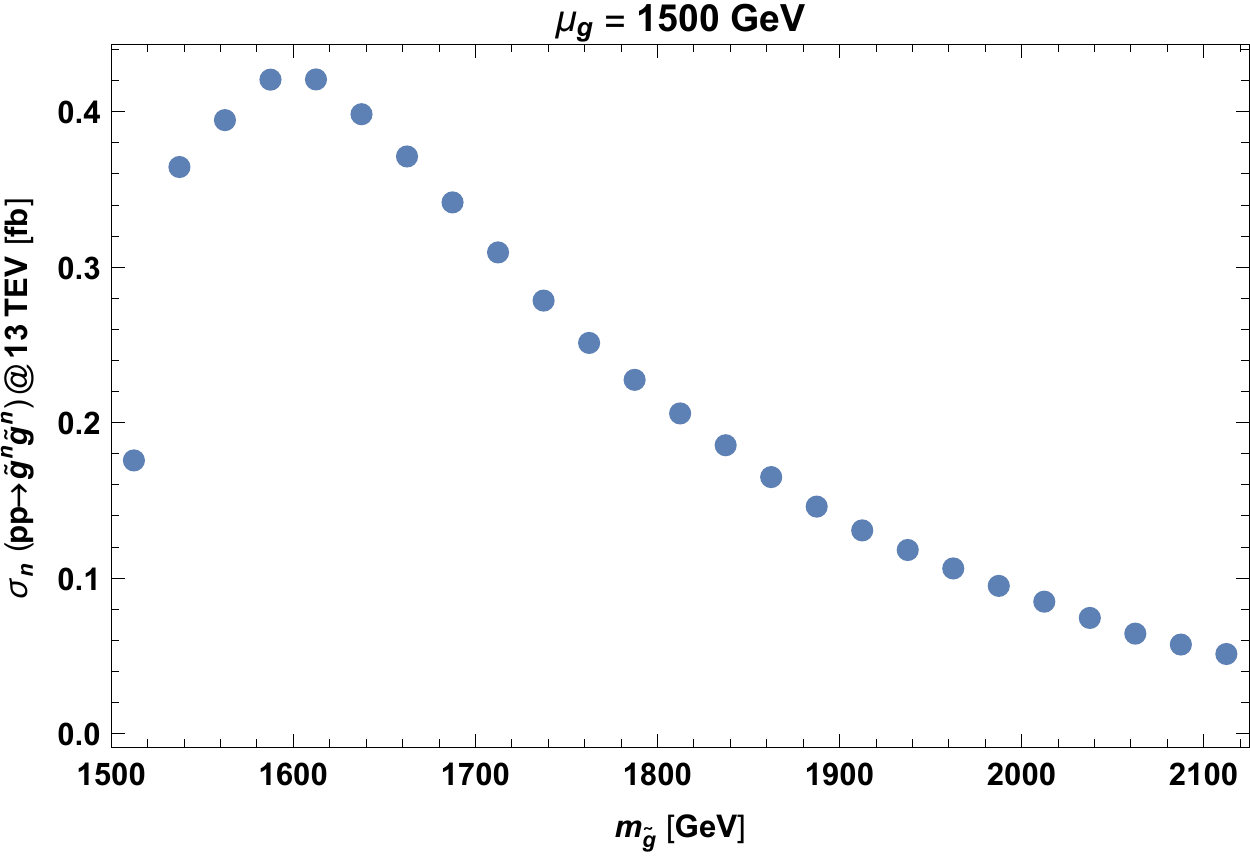}
        \caption{$\sigma_n$ vs $m_n$.}
        \label{fig:xseckk}
    \end{subfigure}
    ~ 
   \caption{Differential cross section of pair producing gluino continuum with $\mu_g=1.5$ TeV at the LHC with 13 TeV. The right panel is a discrete estimate of the left panel, where the first 20 KK modes make up $\gtrsim 85\%$ of the total cross section. }
\end{figure}

With $z_{IR}=0.1$ GeV$^{-1}$, we see in Fig.~\ref{fig:xseckk} that the first 20 KK modes make up $\gtrsim 85\%$ of the total cross section of the chosen benchmark, therefore, we approximate the gluino continuum as the sum of the first 20 KK gluinos. Under these assumptions, we can simulate the production and the decay of each KK mode independently, with the sum of modes mimicing the behavior of a gluino continuum at the LHC. As we shall see, the phenomenology of the gluino continuum is basically determined by two parameters, the threshold $\mu_g (<\mu_f)
$ and the LSP mass, $m_{\chi^0}$.

If $\mu_f\gg \mu_g$, the KK modes that ``constitute" the continuum undergo 3-body decays via off-shell continuum squark. For the highly excited KK gluino, there exists a competition between its decay to a lower gluino KK mode $+$jets and its decay to the LSP $+$ jets. This largely depends on the LSP mass, $m_{\chi^0}$. If $m_{\chi^0}\ll\mu_g$, the entire KK gluino tower decays promptly to $\chi^0+$ jets, because the KK modes' decays amongst themselves are much more suppressed due to a smaller phase space. 
The LHC signature for this scenario will be very similar to that of MSSM gluino with a mass $\gtrsim\mu_g$, hence fairly well constrained by the standard Jets $+$ Missing Transverse Energy (MET) searches \cite{CMS:2019twi}.
If $m_{\chi^0}\lesssim \mu_g$, the higher KK modes can undergo cascade decays via the lower KK modes and the lower KK modes can have a macroscopic lifetime! Displaced vertex searches with multiple jets \cite{Sirunyan:2018pwn} may be sensitive to this special corner of parameter space.


\subsection{Benchmark Study}
If $\mu_f \gtrsim \mu_g$, KK gluinos undergo long cascade decays as discussed in section~\ref{prodanddecay}. To investigate its LHC phenomenology, we perform a detailed study for a gluino continuum with a threshold of $\mu_g=1.5$ TeV accompanied by four species of squarks $(\tilde{u},\tilde{d},\tilde{c},\tilde{s})$ of $\mu_f=1.55$ TeV. We will compare this benchmark with a MSSM gluino of $m_{\tilde{g}}=1.65$ TeV so that the production cross sections of MSSM gluino is roughly the same as the gluino continuum. We will present the result for three different  LSP mass, $0.1, 0.5$, and $1.2$ TeV.\\

We use MadGraph5 \cite{Alwall:2011uj} to generate the parton level events, that is, the production of the first two KK gluinos. We then decay each gluino, step by step according to KK gluino and subsequent squark's branching ratios, examples of which are given in Section~\ref{prodanddecay}. We then pass the final particles to PYTHIA 8 \cite{Sjostrand:2007gs} for showering and hadronization. Finally, the detector simulation is done using DELPHES 3 \cite{deFavereau:2013fsa}.

\begin{figure}
\centering
\begin{subfigure}[b]{0.32\textwidth}
\includegraphics[scale=0.25]{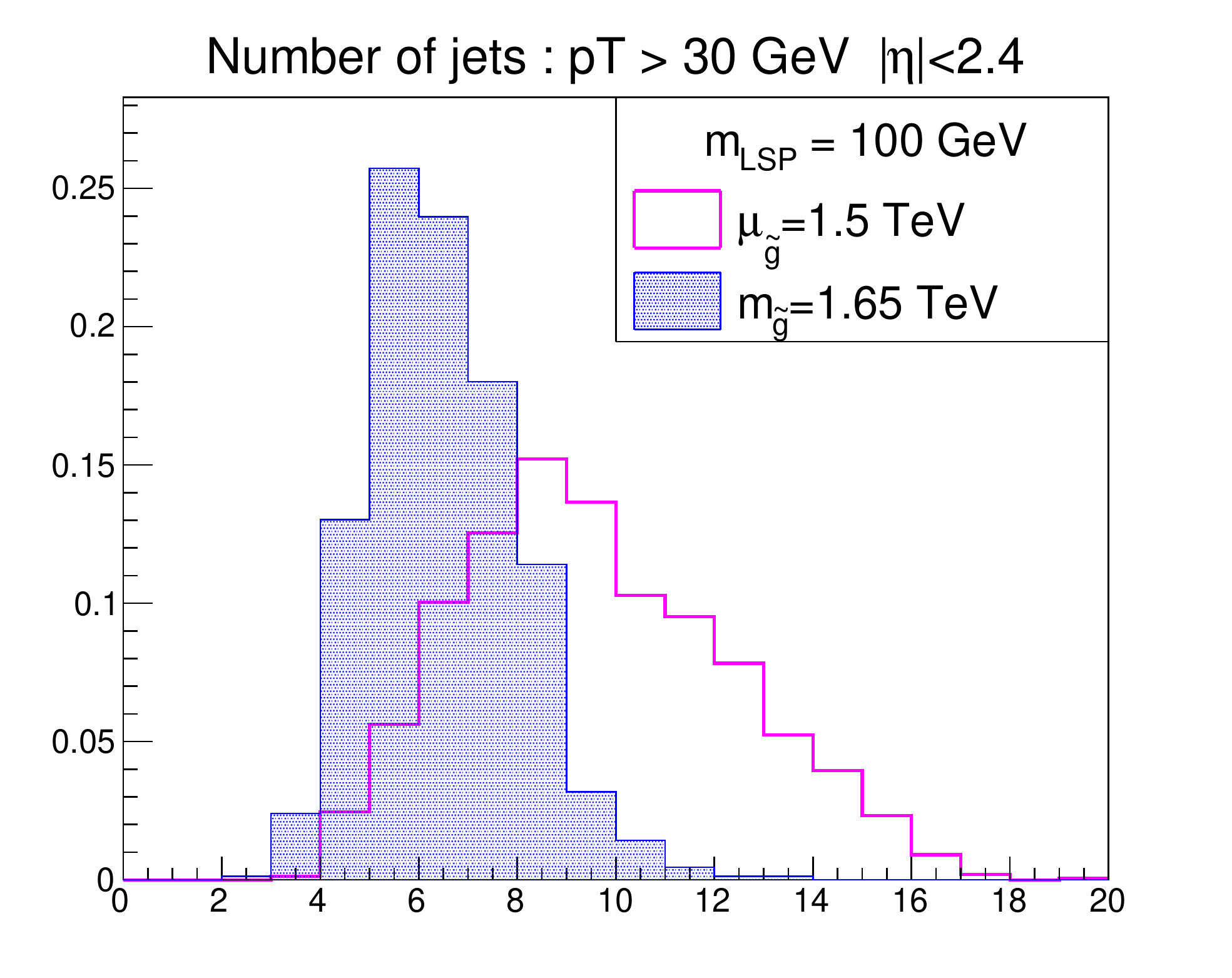}
\caption{
}
\label{nj1}
\end{subfigure}
\begin{subfigure}[b]{0.32\textwidth}
\includegraphics[scale=0.25]{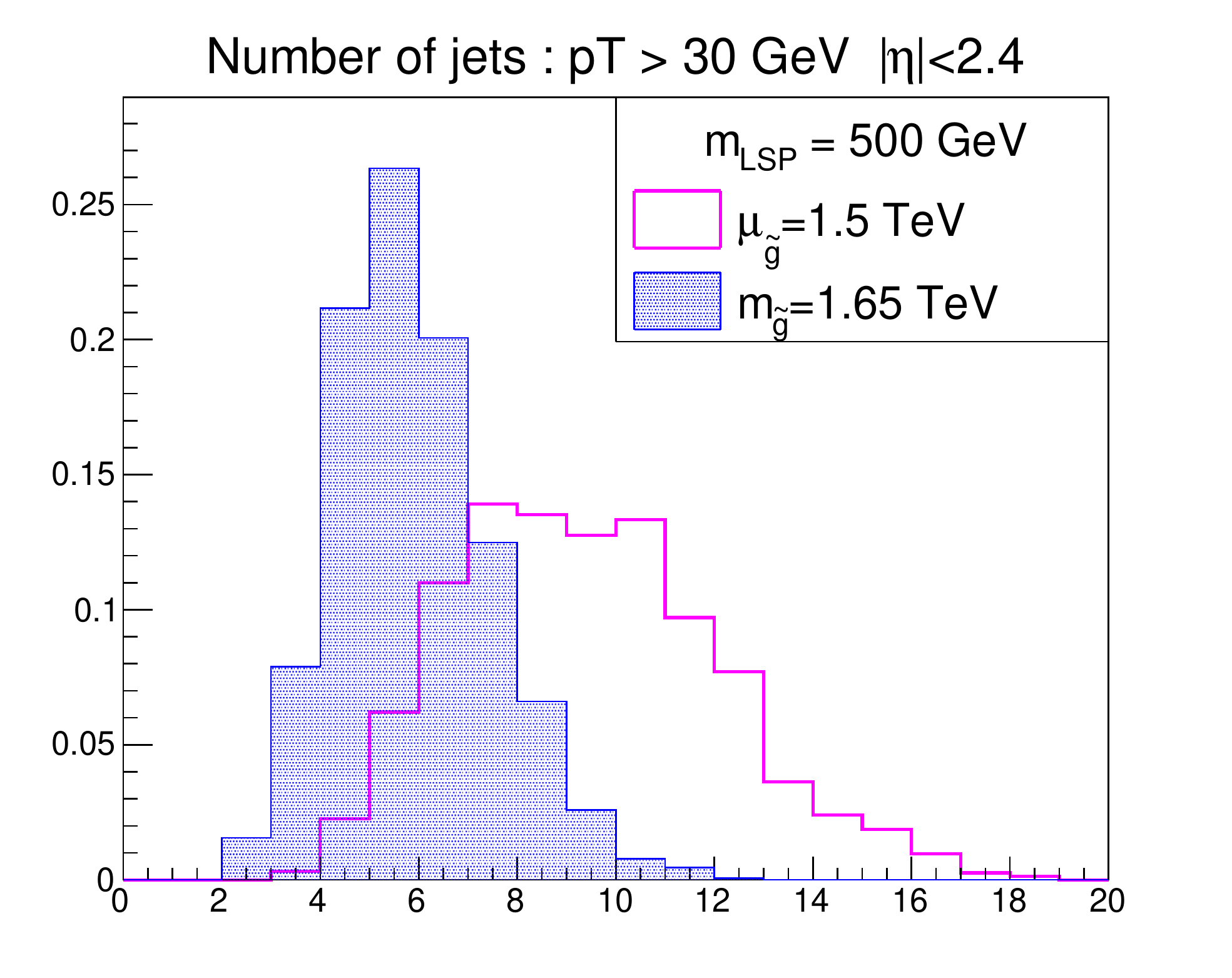}
\caption{
}
\label{nj2}
\end{subfigure}
\begin{subfigure}[b]{0.32\textwidth}
\includegraphics[scale=0.25]{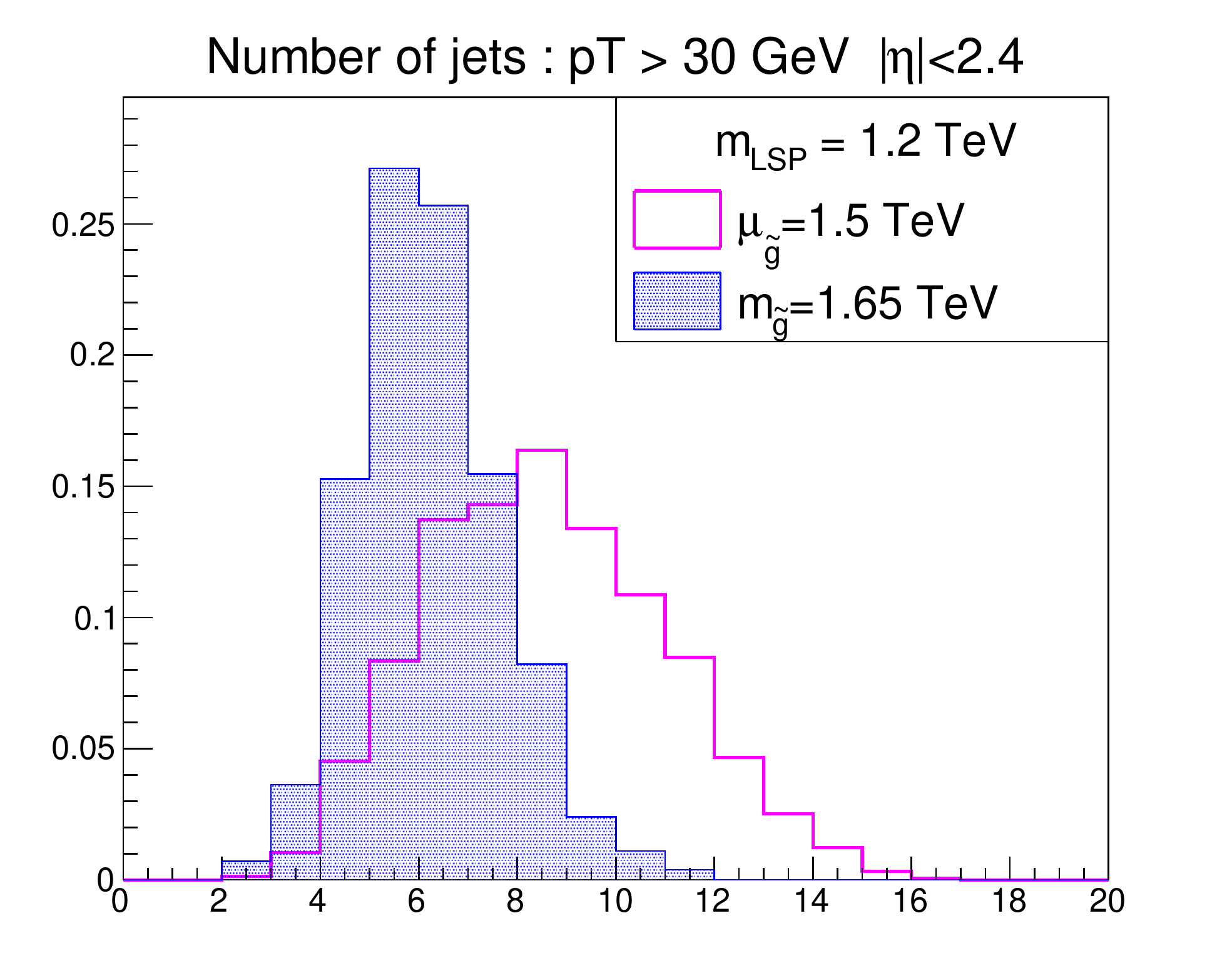}
\caption{
}
\label{nj3}
\end{subfigure}
\begin{subfigure}[b]{0.32\textwidth}
\includegraphics[scale=0.25]{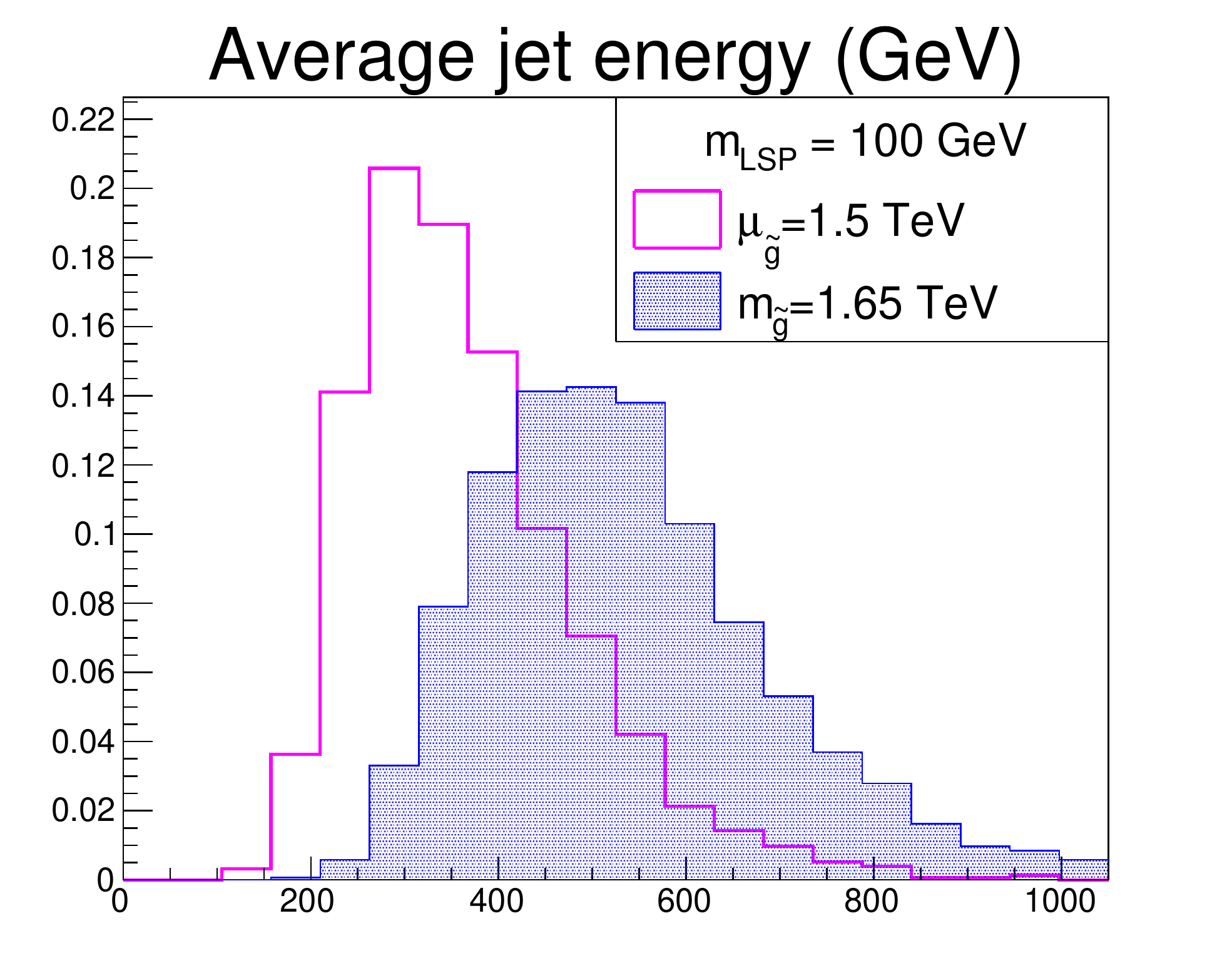}
\caption{
}
\label{aveje1}
\end{subfigure}
\begin{subfigure}[b]{0.32\textwidth}
\includegraphics[scale=0.25]{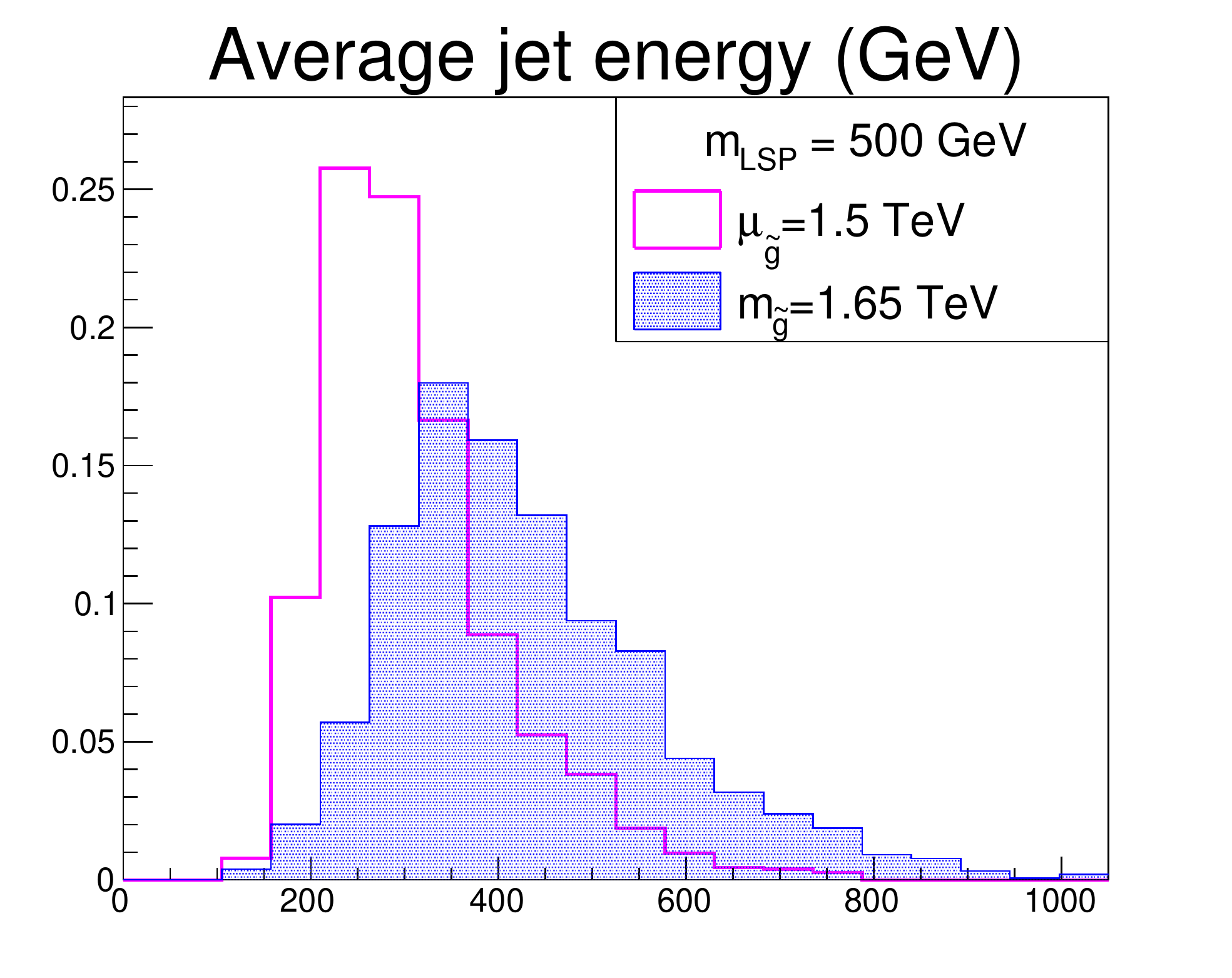}
\caption{
}
\label{aveje2}
\end{subfigure}
\begin{subfigure}[b]{0.32\textwidth}
\includegraphics[scale=0.25]{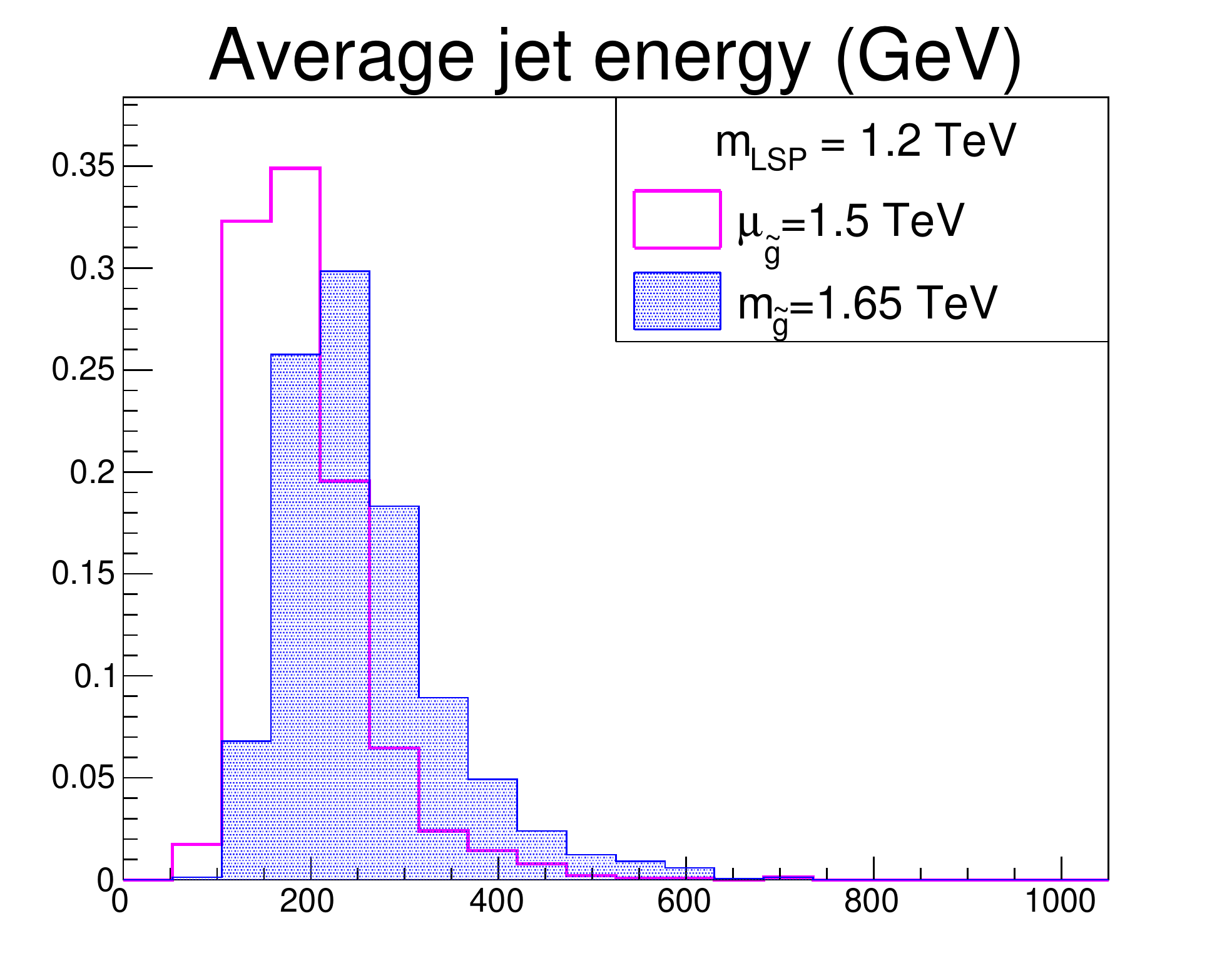}
\caption{
}
\label{aveje3}
\end{subfigure}
\caption{Comparison of MSSM gluino (blue) and gluino continuum (magenta) for different masses of LSP at the LHC with 13 TeV. Distributions are all normalized to 1. The chosen benchmarks are an MSSM gluino with a mass of 1.65 TeV and a gluino continuum with $\mu_g=1.5$ TeV, which have equal production cross sections.}
\end{figure}

\begin{figure}
\centering
\begin{subfigure}[b]{0.32\textwidth}
\includegraphics[scale=0.25]{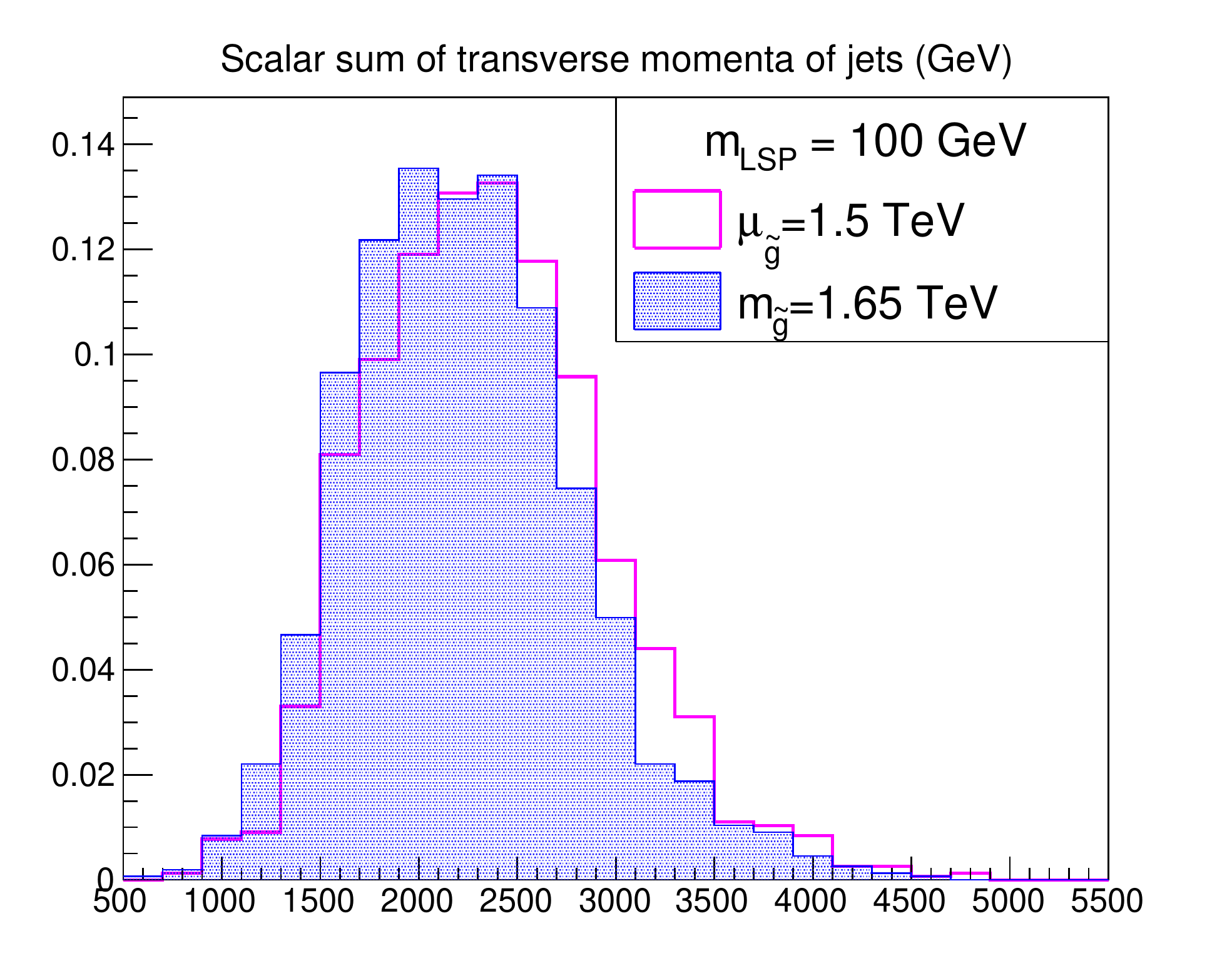}
\caption{
}
\label{ht1}
\end{subfigure}
\begin{subfigure}[b]{0.32\textwidth}
\includegraphics[scale=0.25]{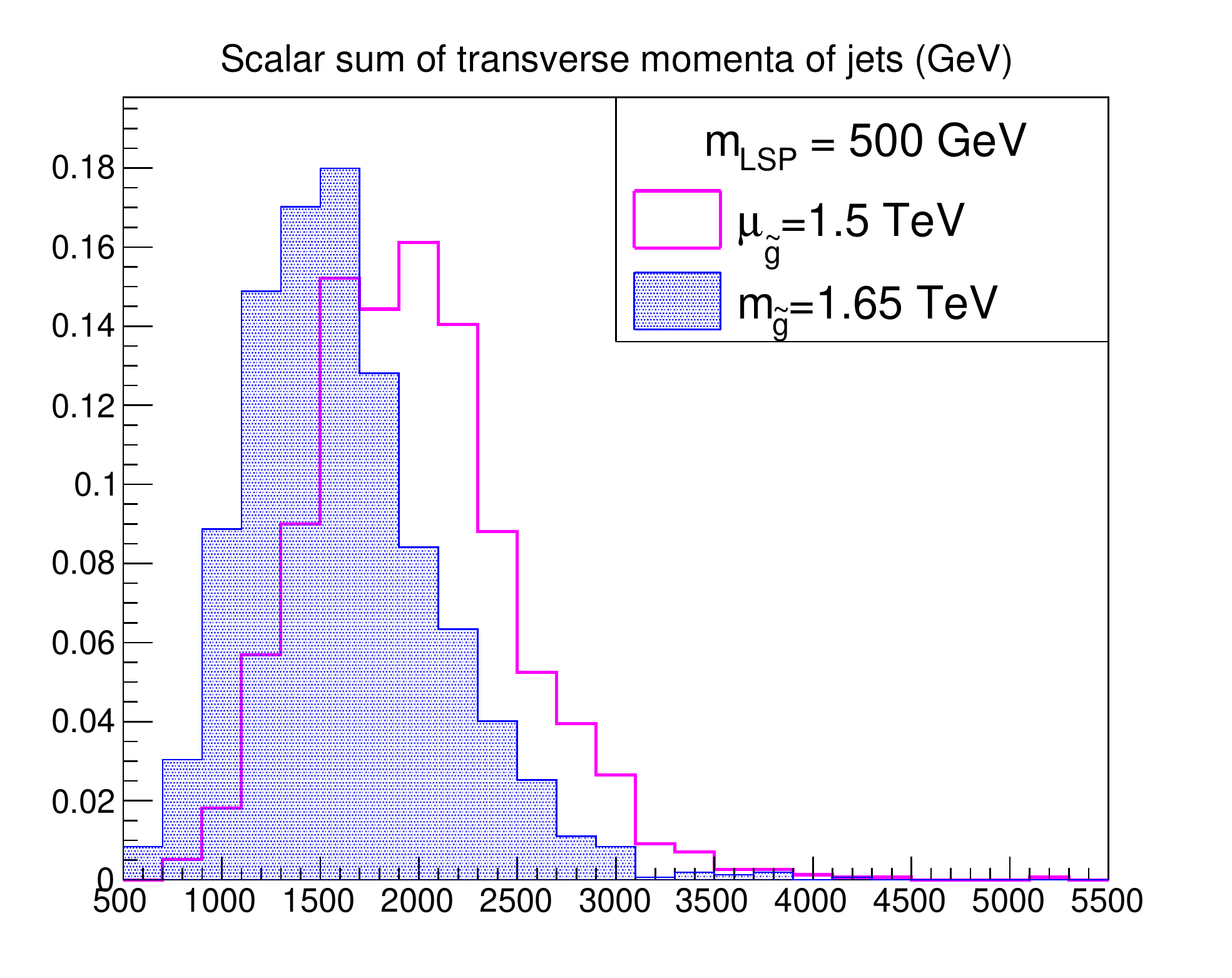}
\caption{
}
\label{ht2}
\end{subfigure}
\begin{subfigure}[b]{0.32\textwidth}
\includegraphics[scale=0.25]{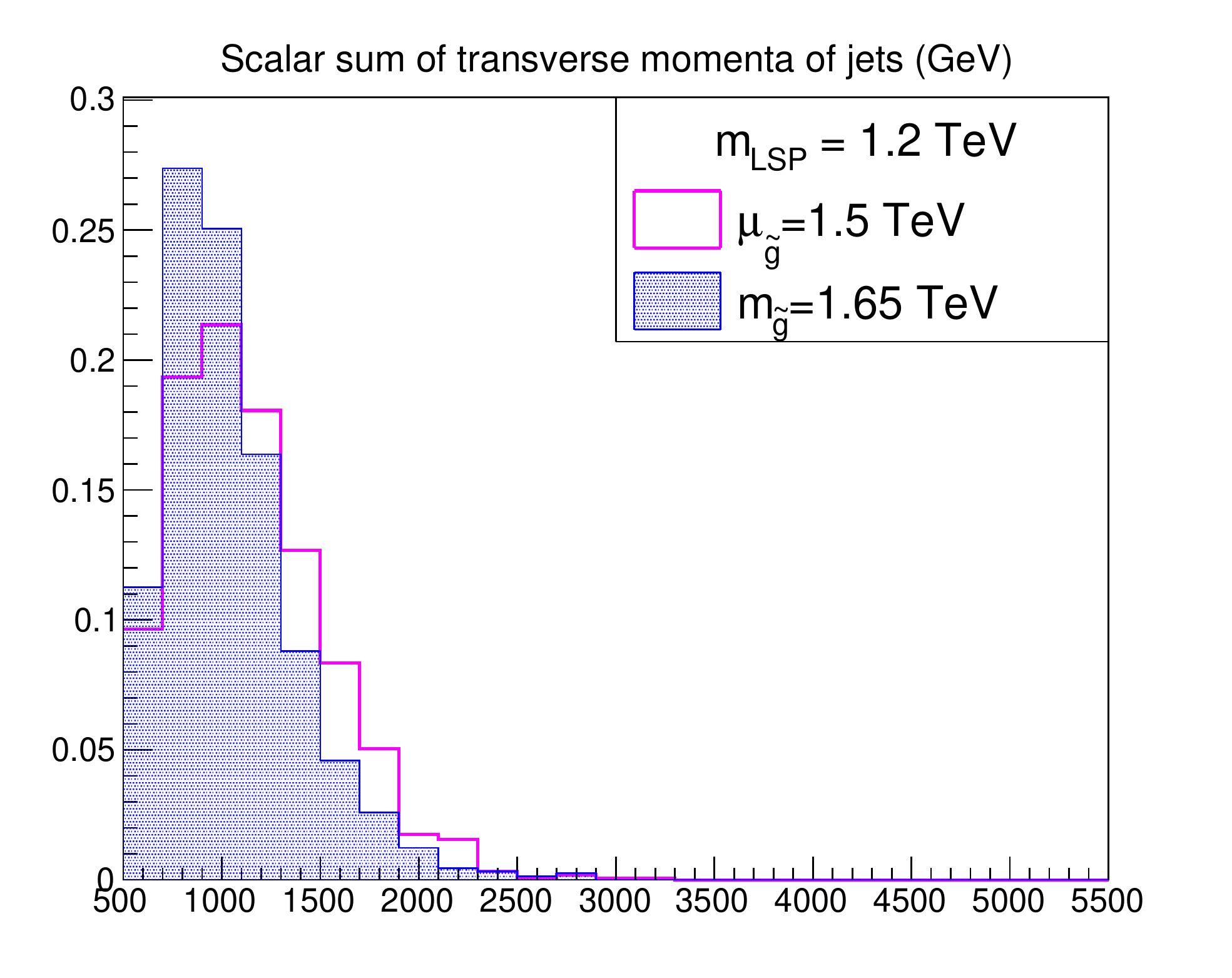}
\caption{
}
\label{ht3}
\end{subfigure}
\begin{subfigure}[b]{0.32\textwidth}
\includegraphics[scale=0.25]{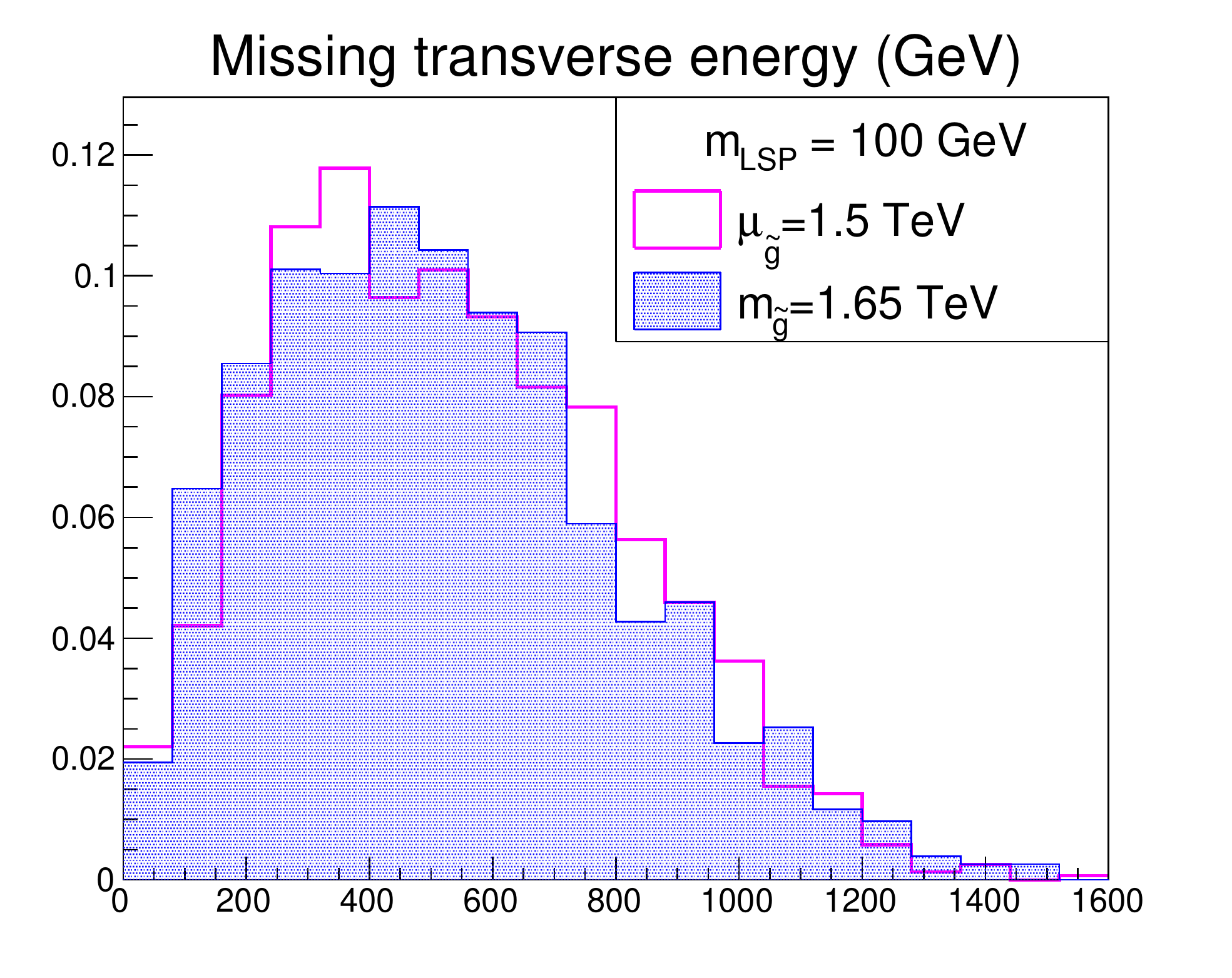}
\caption{
}
\label{met1}
\end{subfigure}
\begin{subfigure}[b]{0.32\textwidth}
\includegraphics[scale=0.25]{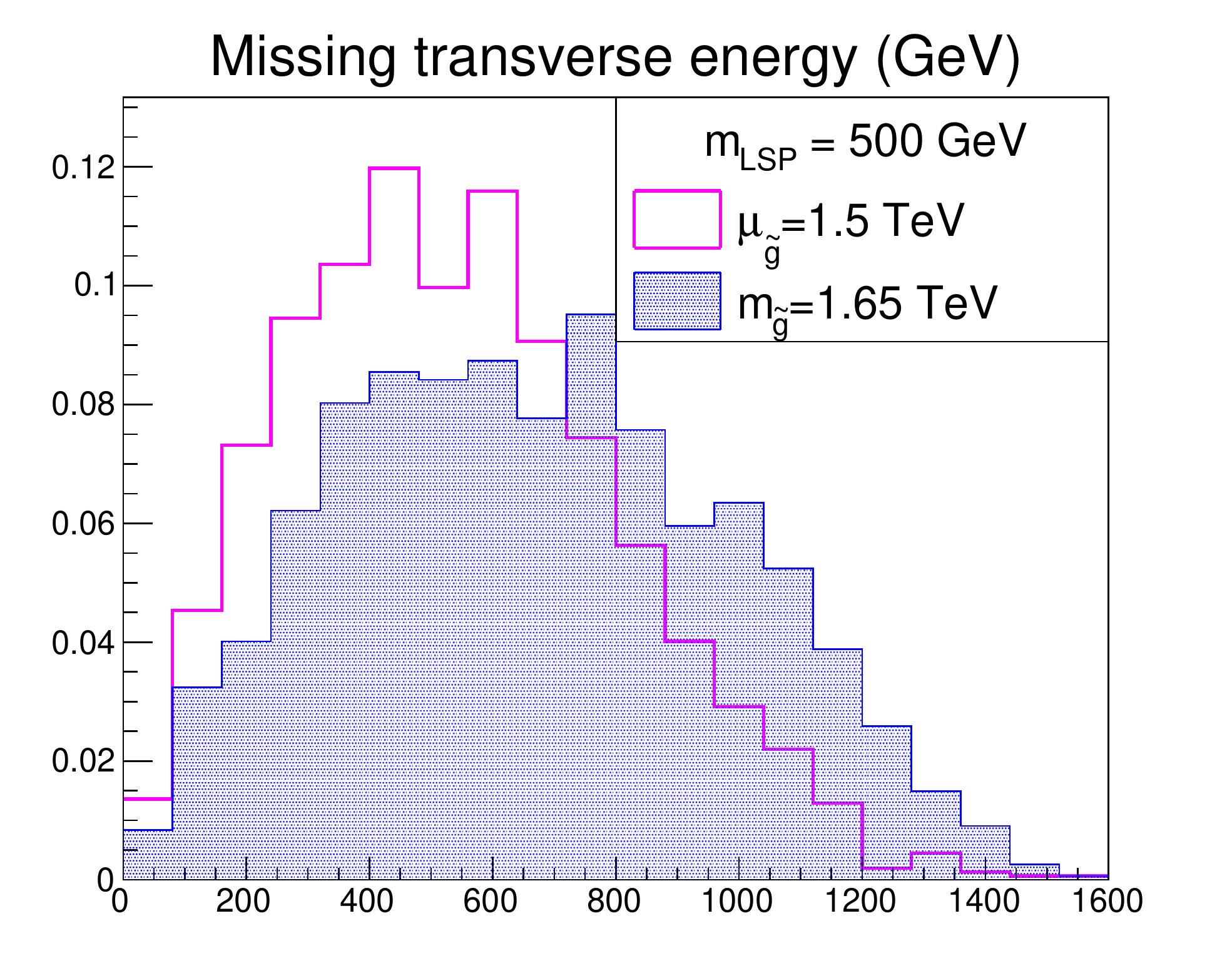}
\caption{
}
\label{met2}
\end{subfigure}
\begin{subfigure}[b]{0.32\textwidth}
\includegraphics[scale=0.25]{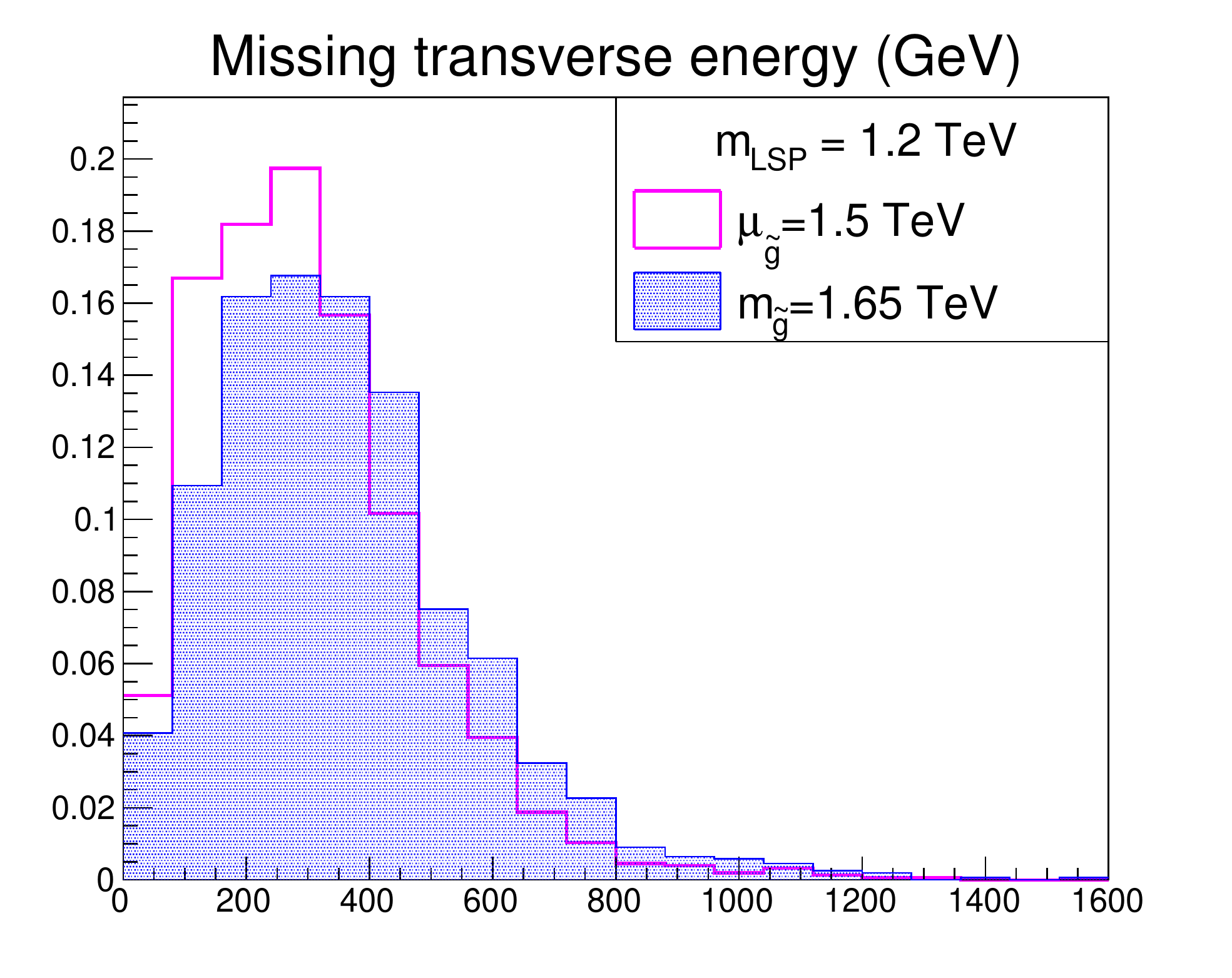}
\caption{
}
\label{met3}
\end{subfigure}
\caption{Comparison of MSSM gluino (blue) and gluino continuum (magenta) for different masses of LSP at the LHC with 13 TeV. Distributions are all normalized to 1. The chosen benchmarks are an MSSM gluino with a mass of 1.65 TeV and a gluino continuum with $\mu_g=1.5$ TeV, which have equal production cross sections.}\label{HTandMET}
\end{figure}

Fig.~\ref{nj1} to \ref{nj3} compare the number of jets with $p_T>30$ GeV and $|\eta|<2.4$ for continuum and MSSM gluinos with three difference choices of LSP mass . 
As expected, the continuum signal typically has more jets due to the long cascade decays. 
Accordingly, as shown in Fig.~\ref{aveje1} to \ref{aveje3}, the average jet energy in the continuum case is smaller. The initial gluino, whether it is an MSSM gluino or a (regulated) KK gluino, before its decay carries similar amount of energy for both cases, so the more jets found in the final state, on average the less energy each of them must have.

We also observe that as the mass of the LSP increases, the difference between the mean jet energy of the continuum case and the MSSM decreases. The jets from the 2-body decays in the cascade of the gluino continuum are far less energetic compared to the jets from the 3-body decay $\tilde{g}\to \tilde{\chi}^0 j j$, which occurs at the end of the decay chain for the gluino continuum. The MSSM gluino, on the other hand, can only decay via the 3-body process, and as the LSP gets heavier, the 3-body jets get less and less energetic, hence their average is closer to the mean jet energy in the continuum case, as can be seen in Figs.~\ref{aveje1} to \ref{aveje3}.
 
Fig.~\ref{ht1} to Fig.~\ref{ht3} show the scalar sum of all the jet $p_T$'s ($H_T$). We see that gluino continuum case has a slightly bigger $H_T$, due to the extra jets from the 2 body decays.
Fig.~\ref{met1} to \ref{met3} show the transverse missing energy (MET) distributions. One can think of it as the vector sum of all the jet $p_T$s in this case. Since there are more jets in the gluino continuum decay, their emission can be more spread out in the transverse plane, yielding a smaller MET.
The effects of a boost in $H_T$ and decrease in MET due to the extra softer jets in the gluino continuum decays are less prominent in the case of either very light ($m_{\chi^0}=100$ GeV) or heavy ($m_{\chi^0}=1200$ GeV) LSP.
This is easy to understand. With a very light LSP, jets from the three body decays are so energetic that the existence of the extra softer jets hardly have an impact on the final state kinematics. 
With a very heavy LSP, the overall jet activity is reduced, hence a fractional increase due to the extra softer jets can hardly be seen.
It is only in the intermediate LSP mass range where we see a detectable difference between the continuum and MSSM gluino decays.

\begin{table}[h]
\caption{Cutflow for the benchmark study, which looks at the $10j0b$ bin from CMS's inclusive $\text{M}_{\text{T}2}$ search using 137fb$^{-1}$ data.}
\centering
\begin{tabular}{ ccccc}
\hline
\hline
 $(\mu_{\tilde{g}},m_{\text{LSP}})$&
  Initial & $H_T>$1500\&$\text{M}_{\text{T}2}>$400 &$10j0b$ &   $\Delta\phi(p_\text{T}^{\text{miss}},j_{1,2,3,4})$\\
GeV&@137fb$^{-1}$&\&$p_{\text{T}}^{\text{miss}}>$30 GeV&& $>0.3$\\
\hline
\hline
(1500,100)& 706& 667 & 121   &86\\
(1500,500)& 706& 584  & 105  &75\\
(1500,1200)& 706& 119  & 27  & 20\\
 \hline
\end{tabular}
\label{table:1}
\end{table}

Due to the existence of extra jets from the long cascade in the gluino continuum decay and the fact that they do not alter the final state kinematics too much compared to MSSM gluino decay, 
the type of gluino continuum decay studied in this section is within the target of traditional SUSY searches. 
For instance, CMS's inclusive MT2 search \cite{CMS:2019twi} has a $10j0b$ bin which would pick up the continuum signal easily for the first two benchmarks with an LSP mass of $0.1$ TeV and $0.5$ TeV (Table.~\ref{table:1}). The fact that this bin does not have any excess with a luminosity of 137$^{-1}$fb rule out these two benchmark points.
The last benchmark with LSP mass of $1.2$ TeV represents the so-called \emph{compressed} type of spectrum which often has little MET. 
There exist search techniques targeting this special type of SUSY spectrum, often including triggering on a hard initial state radiation (ISR) jet and constructing special kinematic variables.

Other than traditional SUSY searches, black-hole-like (BH) searches that are based on looking for an excess of broad spectrum of $H_T$ can be sensitive to a general continuum that couples to the SM sector. However, BH searches look for excesses at very high $H_T$ to avoid QCD backgrounds. For instance, Ref.~\cite{Sirunyan:2017anm} only looked at the $S_T$ (sum of $H_T$ and MET) distribution greater than 2 TeV. This cut on $S_T$ potentially excludes the continuum gluino signal depending on the assumption made of the mass of LSP (Fig.~\ref{HTandMET}).

In conclusion, the gluino continuum decay has the following features in comparison with the MSSM gluino:
\begin{itemize}
	\item significantly more jets
	\item slightly larger $H_T$
	\item slightly smaller MET
\end{itemize}
Therefore, with the LSP assumed to have a particle nature, rather than unparticle, the current run of LHC can be sensitive to gluino continuum decays, but will produce weaker constraints for comparable gluino mass thresholds.

\section{Result and Conclusion}\label{conclusion}
We have investigated the phenomenology of gluino continuum cascade decays at the LHC by examining the production and decay of a regulated gluino continuum, taking care so that physical observables, such as the cross section and the decay probabilities, are regulator ($z_{IR}$) independent. Assuming that the LSP is a particle, we performed a detailed benchmark study and found that the signature of a gluino continuum is broadly similar to that of the MSSM gluino except for very special corners of parameter space, where lighter KK gluinos can start to become long-lived.  At generic points there is an increase in the number of jets and a modest reduction in MET.
This should not be too surprising. Since all the KK modes must decay down the ``ladder"  to the lightest KK gluino, which then generically decays like the usual MSSM gluino decays. The signature is simply the MSSM gluino decay with additional, softer jets emitted while going down the ladder. This picture can change qualitatively if the LSP itself becomes part of a continuum. For example, if the continuum LSP is mainly a bino, the highly excited KK bino can undergo 3-body decays to a lighter KK bino and two jets via off-shell squarks, therefore further depleting the transverse missing energy. This offers new directions to look for SUSY hiding in a continuum.  The methods we have developed here should also be easily transferrable to other models with similar phenomenology, most notably  ``clockwork/linear dilaton" models \cite{Giudice:2017fmj}.

\vspace{24pt}
\textbf{Acknowledgements} 
It is CG's pleasure to thank Y. Bai, H.-C. Cheng, K. C. Kong, and \emph{all} the PPD theory colleagues at Fermilab for useful discussions and encouragement. 
We also thank J.S. Lee and S. Lombardo's input at the initial stage of this work.
This work was supported in part by the DOE under grant DE-SC-0009999.
The work of CG was supported by Fermilab, operated by Fermi Research Alliance, LLC under contract number DE-AC02-07CH11359 with the United States Department of Energy.


\pagebreak
\appendix

\section{Holographic Action}\label{app:holoAction}
According to the  AdS/CFT conjecture, the same physics can be described by  a strongly coupled  CFT or by an  AdS bulk theory. 
Since we are interested in probing the LH bulk fields, we fix the boundary value of the RH superfield $\Phi_c(z_{UV})=\Phi_0$. This can be achieved by adding a boundary action to the original bulk action:
\begin{equation}
\begin{split}
S_{UV}=&-\int d^4x \frac12\left(\frac{R}{z_{UV}}\right)^3\big(\int d^2\theta \Phi(z_{UV})\Phi_0+h.c.\big)\\
=&-\int d^4x\frac12\left(\frac{R}{z_{UV}}\right)^3\big(\phi F_0+F\phi_0+\chi\psi_0+h.c.\big)~.
\end{split}
\end{equation}
Now the UV boundary conditions from varying the fields are modified:
\beq
\delta S_{UV}&=&
\int d^4x \frac12 \left(\frac {R}{z_{UV}}\right)^3 \left[
(F_c-F_0)\delta\phi-\delta F_c\phi+(\phi_c-\phi_0)\delta F
-\delta\phi_c F  \right. \nonumber \\
&&\quad\quad\quad\quad\quad\quad\quad\quad \left. +(\psi-\psi_0)\delta\chi-\chi\delta\psi+h.c. \right]
\eeq
Therefore,
\begin{equation}\label{bc}
\psi(z_{UV})=\psi_0~,\quad \phi_c(z_{UV})=\phi_0~,\quad F_c(z_{UV})=F_0~,
\end{equation}
as desired.
Then one puts the bulk fields on-shell using the equations of motion, which leaves us with only a boundary action:
\begin{equation}
S_{\rm holo}=S_{UV}=-\int d^4x\frac12\left(\frac{R}{z_{UV}}\right)^3\big(\phi F_0+F\phi_0+\chi\psi_0+h.c. \big)~.
\end{equation}
We can think of the boundary fields $\psi_0,\phi_0,F_0$ as the sources coupled to the bulk fields $\chi,F,\phi$, respectively.
From \eqref{bc}, we can normalize the field $\psi(p,z)$ as:
\begin{equation}
\psi(p,z)=\left(\frac z{z_{UV}}\right)^{3/2}\frac{f_R(p,z)}{f_R(p,z_{UV})}\psi_0(p)~.
\end{equation}
Comparing with \eqref{rhfield}, we see that
\begin{equation}
\psi_4(p)\equiv  \frac{\psi_0(p)}{f_R(p,z_{UV})}~.
\end{equation}
Using the coupled equations of motion of  the Dirac fields $\psi_4$ and $\chi_4$ we get
\begin{equation}
\chi(p,z)=\left(\frac z{z_{UV}}\right)^{3/2}\frac{f_L(p,z)}{f_R(p,z_{UV})}\frac{p_{\mu}\sigma^{\mu}}{p}\bar{\psi}_0(p)~.
\end{equation}
Therefore, the fermionic holographic action is given by
\begin{equation}\label{holoaction}
S_{\rm holo}[\psi_0]=-\int \frac{d^4 p}{(2\pi)^4}\left(\frac{R}{z_{UV}}\right)^3
\frac{f_L(p,z_{UV})}{f_R(p,z_{UV})}\frac{p_{\mu}\sigma^{\mu}}{p}\bar{\psi}_0(p)\psi_0(-p)~.
\end{equation}
In position space we have:
\beq
S_{\rm holo}[\psi_0]&=&-\int d^4 x \int d^4 y \psi_0(y)\Delta^{-1}_\psi (x-y)\bar{\psi}_0(x)~,\\
\Delta^{-1}_\psi (x-y)&=&  \int \frac{d^4 p}{(2\pi)^4}e^{-ip(x-y)}\left(\frac{R}{z_{UV}}\right)^3
\frac{f_L(p,z_{UV})}{f_R(p,z_{UV})}\frac{p_{\mu}\sigma^{\mu}}{p}~.
\eeq
So we see that this action is indeed non-local. 

To interpret these results, we can think of putting fields on-shell as equivalent to integrating them out in the path integral. Schematically,
\begin{equation}
\mathcal{Z}[\psi_0]=e^{iS_{\rm holo}[\psi_0]}=\int d[\mathcal{O}_{\chi}] e^{iS[\mathcal{O}_{\chi}]+i\int \mathcal{O}_{\chi}\psi_0}~.
\end{equation}
Therefore, one can think of the equation above as a generating functional of the 4D CFT field $\mathcal{O}_{\chi}$ with $\psi_0$ acting as a source. This allows us to derive the correlation function between $\mathcal{O}_{\chi}$ by taking functional derivatives of $\ln\mathcal{Z}[\psi_0]$ with respect $i\psi_0$. We are particularly interested in the two point correlation function 
\begin{equation}
\langle \mathcal{O}_{\bar{\chi}}(x)\mathcal{O}_{\chi}(y)\rangle=\frac{\delta S_{\rm holo}}{\delta \bar{\psi}(x)\delta \psi(y)}= \Delta^{-1}_\psi (x-y)~.
\end{equation}
From general knowledge of QFT, we can insert a complete set of states into the correlator, 
\begin{equation}\label{2pt}
\langle \mathcal{O}_{\bar{\chi}}(x)\mathcal{O}_{\chi}(y)
\rangle= \int_{FC} \frac{d^4p}{(2\pi)^4} e^{-ip(x-y)}\sum_{\lambda}|\langle \mathcal{O}_{\chi}|\lambda_0\rangle|^2~,
\end{equation}
where $FC$ stands for forward light cone, which only allows physical states. We see that the spectral density of this theory is proportional to the kinetic function:
\begin{equation}
\Sigma_{\psi}(p)=\left(\frac{R}{z_{UV}}\right)^3
\frac{f_L(p,z_{UV})}{f_R(p,z_{UV})}\frac{p_{\mu}\sigma^{\mu}}{p}~.
\end{equation}
We will see that this agrees with the spectrum in the KK picture when the IR regulating brane is taken to infinity.
From \cite{08terning2}, one can obtain an effective unparticle action for the LH CFT fields $\mathcal{O}_{\chi}$ by applying a Legendre transformation to the holographic action \eqref{holoaction}. It can be shown that the propagator for the unparticle $\mathcal{O}_{\chi}$ is proportional to $\Sigma_{\psi}$.

The kinetic functions for the scalars are 
\begin{align}
\Sigma_{F}(p) =&\left(\frac{R}{z_{UV}}\right)^3
\frac{f_L(p,z_{UV})}{f_R(p,z_{UV})}\frac1p~,
\\
\Sigma_{\phi}(p) =&\left(\frac{R}{z_{UV}}\right)^3
\frac{f_L(p,z_{UV})}{f_R(p,z_{UV})} p~,
\end{align}
which are related to the kinetic functions of the  fermions by SUSY. 
The solutions for $f_L$ and $f_R$
are Whittaker functions  given by
\begin{equation}
\begin{split}
M(k,m,z)=&z^{m+1/2}e^{-z/2}\sum_{n=0}^{\infty}\frac{(m-k+1/2)_n}{n!(2m+1)_n}z^n~,\\
W(k,m,z)=&\frac{\Gamma(-2m)}{\Gamma(1/2-m-k)}M(k,m,z)+\frac{\Gamma(2m)}{\Gamma(1/2+m-k)}M(k,-m,z)~,\\\mbox{where}\,\,\, (x)_n\equiv&\frac{\Gamma(x+n)}{\Gamma(x)}~.
\end{split}
\end{equation}
We adopt outgoing wave boundary conditions, thus excluding the Whittaker function of the first kind, $M(\kappa,m,z)$, and take $z_{UV}=R$. Therefore,
\begin{equation}
\Sigma_{F}(p)=
\frac{W(\kappa,\frac12+c,2\sqrt{\mu^2-p^2}z_{UV})}{W(\kappa,-\frac12+c,2\sqrt{\mu^2-p^2}z_{UV})}\frac{\sqrt{\mu^2-p^2}+\mu}{p^2}~.
\end{equation}


\section{The KK Tower and the Continuum Limit}\label{app:KKvsCon}
We see that in \eqref{2pt}, one can insert a complete set of states to get the spectral density for the fields. In the holographic picture, we get the kinetic function as a result. 
In the KK picture (with an IR regulator brane), the complete set of states now contain all the KK modes above the mass gap:

\begin{equation}
\begin{split}
\langle \mathcal{O}_{\bar{\chi}}(x)\  \mathcal{O}_{\chi}(y)
\rangle=& \int_{FC}\sum_{n} \frac{d^3p}{(2\pi)^32E_n} e^{-ip(x-y)}|\langle \mathcal{O}_{\chi}|m_n\rangle|^2\\
=& \int_{FC}\sum_{n} \frac{d^4p}{(2\pi)^4} e^{-ip(x-y)}\frac{i}{p^2-m_n^2+i\epsilon}| f_L(m_n,z_{UV})|^2~.
\end{split}
\end{equation}
Therefore, we expect that for $p>\mu$:
\begin{equation}\label{kkprop}
\mbox{Im}P_{KK}=\mbox{Im}\sum_n \frac{1}{p^2-m_n^2+i\epsilon}|f_L(m_n,z_{UV})|^2\xrightarrow{z_{IR}\to\infty} \mbox{Im}\Sigma_{F} 
\end{equation}
for the scalar case. 

\begin{figure}
\centering
\includegraphics{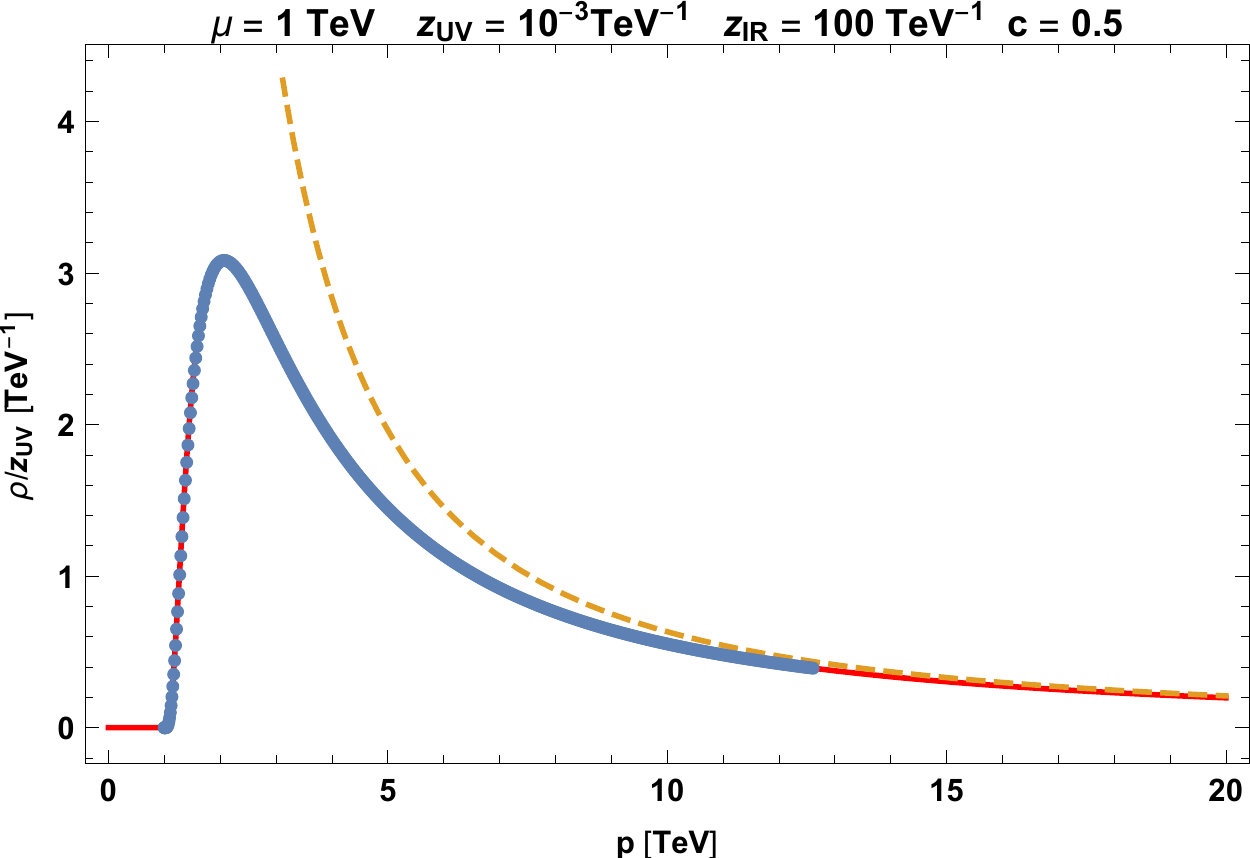}
\caption{Blue dots are calculated from the KK picture in \eqref{kk_result}. The red curve is the continuum spectral function given by $\text{Im}\,\Sigma_F$. The yellow dashed line is an approximation to the spectral function valid only when $\mu\ll p\ll 1/z_{UV}$.}
\label{spectral}
\end{figure}

Taking $z_{IR}\to\infty$ in the KK picture, the masses of the KK modes become more and more closely spaced and the sum in \eqref{kkprop} becomes an integral. One should also  include the Jacobian factor when taking such a limit. From \eqref{kkmass}, we have
\begin{equation}
1=\Delta n = \frac{z_{IR}}{\pi}\frac{m_n\delta m_n}{\sqrt{m_n^2-\mu^2}}= \Big(\frac{z_{IR}}{\pi}\Big)^2\frac{m_n\delta m_n}{\frac14+n}~.
\end{equation}
Therefore,
\begin{equation}
P_{KK}\to \int d m_n\Big(\frac{z_{IR}}{\pi}\Big)^2\frac{m_n}{\frac{1}{4}+n}
\frac{|f_L(m_n,z_{UV})|^2}{p^2-m_n^2+i\epsilon}~.
\end{equation}
Cauchy's integral formula can be used to obtain the imaginary part of the integral above, namely
\begin{equation}\label{kk_result}
\mbox{Im}P_{KK}\xrightarrow{z_{IR}\to\infty}\frac{z^2_{IR}}{2\pi\Big(\frac{1}{4}+n\Big)} |f_L(p,z_{UV})|^2~,\,\,\,  \forall p=m_n~.
\end{equation}
Fig.~\ref{spectral} compares the spectral density function calculated in both KK and continuum picture for a chosen benchmark. An approximation to the spectral function valid only when $\mu\ll p\ll 1/z_{UV}$ is also plotted. Its analytic form is given by \cite{11terning}:
\begin{equation}
\frac{\pi}2 \frac1 {p^2z_{UV}\Bigg(\frac{\pi^2}4+\Big(\gamma_E+\log\Big[\frac{z_{UV}}2\sqrt{p^2-\mu^2}\Big]\Big)^2\Bigg)}~.
\end{equation}

\pagebreak

\end{document}